\documentclass[5p,number,sort&compress]{elsarticle}

\usepackage[T1]{fontenc}
\usepackage[utf8]{inputenc}
\usepackage{xspace}
\usepackage{graphicx}
\usepackage[separate-uncertainty=true,range-phrase=--,range-units=single,product-units=power,group-separator={,}]{siunitx}
\usepackage[super]{nth}
\usepackage{amsmath}
\usepackage{amssymb}
\usepackage{booktabs}
\usepackage{textgreek}

\newcommand{\xcalibur}{\textit{X-Calibur}\xspace}
\newcommand{\xlcalibur}{\textit{XL-Calibur}\xspace}
\newcommand{\infocus}{InFOC{\textmu}S\xspace}

\DeclareSIUnit\gram{g}

\usepackage{caption}
\journal{Astroparticle Physics}

\begin{document}

\begin{frontmatter}

\title{Performance of the \xcalibur Hard X-Ray Polarimetry Mission during its 2018/19 Long-Duration Balloon Flight}
\author[1,1b]{Q. Abarr}
\author[2]{B. Beheshtipour}
\author[3]{M. Beilicke}
\author[1]{R. Bose}
\author[1]{D. Braun}
\author[4]{G. de Geronimo}
\author[1]{P. Dowkontt}
\author[1]{M. Errando}
\author[5]{T. Gadson}
\author[6]{V. Guarino}
\author[5]{S. Heatwole}
\author[1,1b]{M. Hossen}
\author[7,8]{N. Iyer}
\author[9]{F. Kislat\corref{cor}}
\ead{fabian.kislat@unh.edu}
\author[7,8]{M. Kiss}
\author[10]{T. Kitaguchi}
\author[1,1b,1c]{H. Krawczynski}
\author[5]{J. Lanzi}
\author[11]{S. Li}
\author[1,1b]{L. Lisalda}
\author[12]{T. Okajima}
\author[7,8]{M. Pearce}
\author[5]{Z. Peterson}
\author[1]{L. Press}
\author[1,1b]{B. Rauch}
\author[1]{G. Simburger}
\author[5]{D. Stuchlik}
\author[13]{H. Takahashi}
\author[1]{J. Tang}
\author[13]{N. Uchida}
\author[1,1b]{A. West}

\cortext[cor]{Corresponding authors}

\address[1]{Washington University in St. Louis, 1 Brookings Dr., CB 1105, St. Louis, MO 63130, USA}
\address[1b]{McDonnell Center for the Space Sciences at Washington University in St. Louis}
\address[1c]{Quantum Sensor Center at Washington University in St. Louis}
\address[2]{Formerly: Washington University in St. Louis, now at: Max Planck Institute for Gravitational Physics (Albert Einstein Institute), Leibniz Universität Hannover, Germany, formerly: Washington University in St. Louis, 1 Brookings Dr., CB 1105, St. Louis, MO 63130, USA}
\address[3]{Former affiliation: Washington University in St. Louis, 1 Brookings Dr., CB 1105, St. Louis, MO 63130, USA}
\address[4]{DG Circuits, 30 Pine Rd., Syosset, NY 11791, USA}
\address[5]{NASA Wallops Flight Facility, 32400 Fulton St., Wallops Island, VA 23337, USA}
\address[6]{Guarino Engineering, 1134 S Scoville Ave., Oak Park, IL 60304, USA}
\address[7]{KTH Royal Institute of Technology, Department of Physics, SE-106 91 Stockholm, Sweden}
\address[8]{The Oskar Klein Centre for Cosmoparticle Physics, AlbaNova University Centre, SE-106 91 Stockholm, Sweden}
\address[9]{University of New Hampshire, Department of Physics \& Astronomy and Space Science Center, 8 College Rd., Durham, NH 03824, USA}
\address[10]{RIKEN Nishina Center, 2-1 Hirosawa, Wako, Saitama 351-0198, Japan}
\address[11]{Brookhaven National Laboratory, 98 Rochester St., Upton, NY 11973, USA}
\address[12]{NASA's Goddard Space Flight Center, Greenbelt, MD 20771, USA}
\address[13]{Hiroshima University, Department of Physical Science, 1-3-1, Kagamiyama, Higashi-Hiroshima, 739-8526, Japan}

\begin{abstract}
  \xcalibur is a balloon-borne telescope that measures the polarization of high-energy X-rays in the \SIrange{15}{50}{keV} energy range.
  The instrument makes use of the fact that X-rays scatter preferentially perpendicular to the polarization direction.
  A beryllium scattering element surrounded by pixellated CZT detectors is located at the focal point of the \infocus hard X-ray mirror.
  The instrument was launched for a long-duration balloon (LDB) flight from McMurdo (Antarctica) on December 29, 2018, and obtained the first constraints of the hard X-ray polarization of an accretion-powered pulsar.
  Here, we describe the characterization and calibration of the instrument on the ground and its performance during the flight, as well as simulations of particle backgrounds and a comparison to measured rates. 
  The pointing system and polarimeter achieved the excellent projected performance.
  The energy detection threshold for the anticoincidence system was found to be higher than expected and it exhibited unanticipated dead time.
  Both issues will be remedied for future flights.
  Overall, the mission performance was nominal, and results will inform the design of the follow-up mission \xlcalibur, which is scheduled to be launched in summer 2022. 
\end{abstract}


\end{frontmatter}

\section{Introduction}
The polarization of X-rays carries geometrical information about the innermost regions of compact astrophysical objects that are too small to be spatially resolved at any wavelength with current instruments~\cite{krawczynski_etal_2011}.
Probing the structure of black hole accretion disks allows us to study the behavior of matter in strong gravitational fields, and the strongest magnetic fields in the Universe are found in the magnetosphere of neutron stars.
Hence, studying these systems provides tests of fundamental physical principles, such as the theory of General Relativity, and Quantum Electro-Dynamics.

The scientific potential of X-ray polarimetry has long been recognized, but technical challenges have limited the progress.
Since the flight of the first satellite-based X-ray polarimeter on board OSO-8 in the 1970s, only a small number of polarization measurements with purpose-built and calibrated polarimeters have been made.
The most thoroughly studied object in polarized X-rays is the Crab pulsar and nebula.
OSO-8 found a roughly \SI{20}{\percent} polarization of the \SIlist{2.6;5.2}{keV} emission from the Crab pulsar wind nebula~\cite{weisskopf_etal_1978a}, and a polarization angle in agreement with optical polarization measurements.
The balloon-borne hard X-ray polarimeter, PoGO+, measured a polarization fraction of \SI{20.9+-5.0}{\percent} for phase-integrated emission and $(17.4^{+8.6}_{-9.3})\,\%$ for off-pulse emission, in the \SIrange{20}{160}{keV} band~\cite{chauvin_etal_2017}.
The gamma-ray burst polarimeter POLAR found a similar polarization fraction of $14_{-10}^{+15}\,\%$ for phase-averaged emission in the \SIrange{50}{500}{keV} energy range, but unlike PoGO+ excluded the nebula from the measurement~\cite{li_etal_2022}.
The Soft Gamma-ray Detector (SGD) on board the Hitomi satellite found a phase-integrated polarization fraction of \SI{22.1+-10.6}{\percent}, in agreement with the PoGO+ result~\cite{aharonian_etal_2018}.
Recently, the cubesat-based soft X-ray polarimeter PolarLight found a time-averaged \SIrange{3.0}{4.5}{keV} polarization fraction of $(15.3^{+3.1}_{-3.0})\,\%$, but also found that the on-pulse polarization fraction rapidly decreased from $(28.8^{+7.1}_{-7.3})\,\%$ to $(10.1^{+4.7}_{-5.1})\,\%$ following a pulsar glitch, recovering after about 100 days~\cite{feng_etal_2020}.
Additional polarization measurements of the Crab have been reported by AstroSat CZTI~\cite{vadawale_etal_2018}, INTEGRAL IBIS~\cite{forot_etal_2008}, and INTEGRAL SPI~\cite{dean_etal_2008,chauvin_etal_2013,jourdain_roques_2019}.
However, these instruments were not calibrated for polarization measurements resulting in large systematic uncertainties.

In addition to the observations of the Crab, OSO-8 found some evidence for low levels of polarization in a handful of other sources~\cite{weisskopf_etal_1977,weisskopf_etal_1978b,long_etal_1979,long_etal_1980}.
PoGO+ set an upper limit on the polarization from the accreting stellar-mass black hole Cyg X-1, lending weight to the presence of an extended accretion corona~\cite{chauvin_etal_2018b}.

\xcalibur is a balloon-borne hard X-ray polarimeter operating in the \SIrange{15}{50}{keV} energy range.
Similar to PoGO+ and AstroSat CZTI, it makes use of the fact that photons preferentially scatter perpendicular to their polarization direction resulting in a sinusoidal modulation of the azimuthal distribution of scattered photons.
However, unlike the former two, \xcalibur utilizes focusing optics, which allows the use of a compact detector assembly while maintaining a large effective area, thus significantly increasing the signal-to-background ratio.
The design and optimization of \xcalibur~\cite{guo_etal_aph_2013, kislat_etal_jatis_2018}, detector calibration and beam test~\cite{beilicke_etal_jai_2014}, and its telescope structure~\cite{kislat_etal_jai_2017} have been described in previous papers.

\begin{figure}
  \centering
  \includegraphics[width=\linewidth]{trajectory} \\[1ex]
  \includegraphics[width=\linewidth]{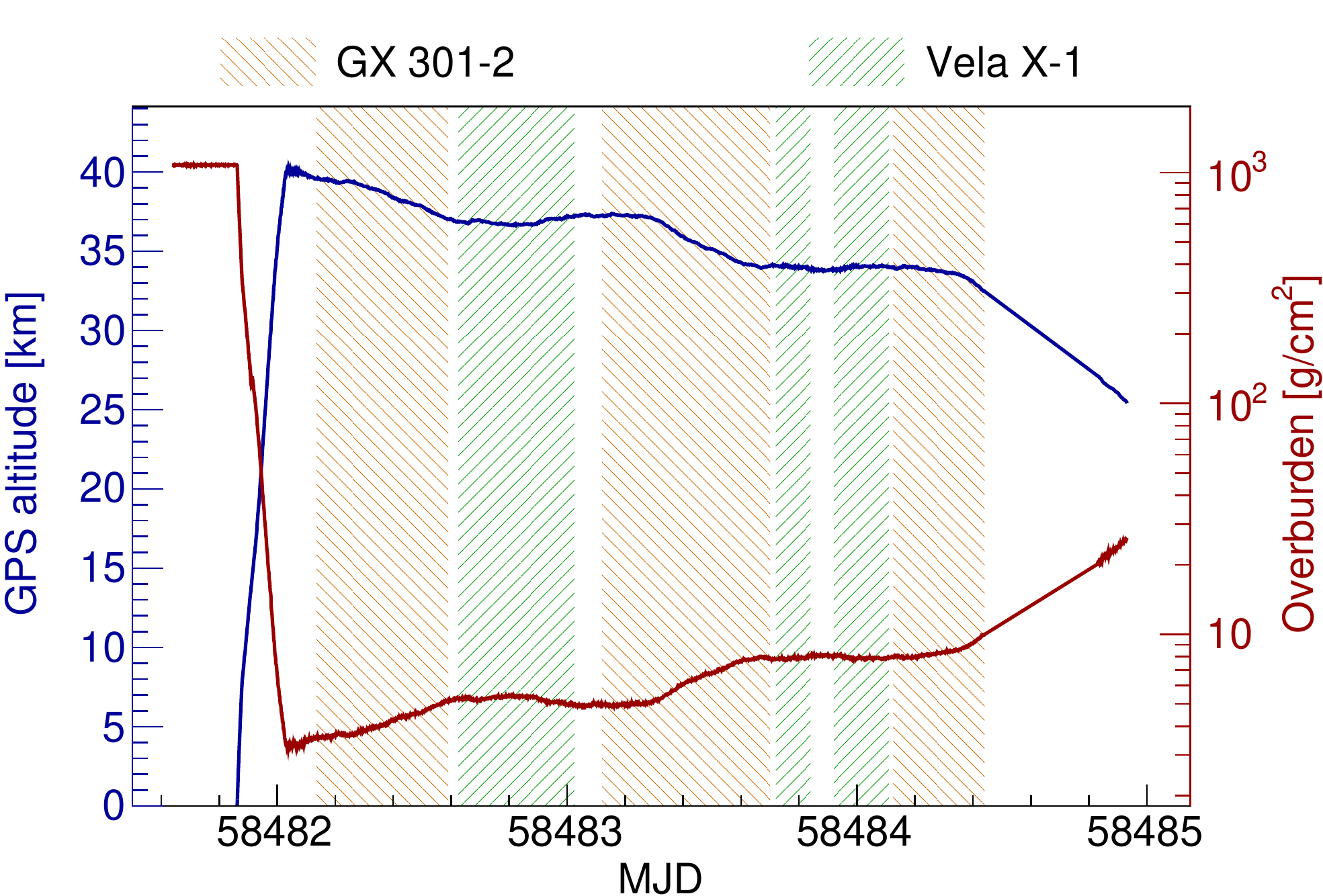}
  \caption{%
    Trajectory (\emph{top}) and flight altitude and atmospheric overburden (\emph{bottom}) of the Antarctic LDB flight of \xcalibur.
    Map: Landsat Image Mosaic of Antarctica team / via Wikipedia (public domain).
    }
  \label{fig:trajectory}
\end{figure}

After test flights from Ft.\ Sumner, NM, in 2013, 2014, and 2016, \xcalibur was launched for a long-duration balloon flight from McMurdo, Antarctica, on December 29, 2018.
Due to a leak in the balloon, the flight had to be terminated less than three days after launch.
Trajectory and flight altitude are shown in Fig.~\ref{fig:trajectory}.
Detector, X-ray mirror, and other important components have been recovered from the remote landing site.
During this flight, X-Calibur performed the first hard X-ray polarization measurement of an accreting X-ray pulsar, GX 301--2.
The science results and some aspects of the data analysis 
have been described in~\cite{abarr_etal_2020}.
Observations of a second object, Vela X-1, were short and performed at a relatively low balloon altitude, so that no excess signal was detected.
Here, we report on the performance of the instrument during this flight, as well as signal and background simulations.

The remainder of the paper is structured as follows:
In Section~\ref{sec:payload} we describe the \xcalibur balloon payload and detector, and discuss the overall performance of the payload.
A detailed account of the detector performance and its evaluation is given in Section~\ref{sec:performance}.
Section~\ref{sec:background} discusses the measured experimental background in comparison with Monte Carlo simulations.
The detector response to incident photons based on a combination of ground calibration and simulations is discussed in Section~\ref{sec:response}.
We close with a summary and an outlook towards the follow-up mission \xlcalibur in Section~\ref{sec:summary}.

\section{Payload Description and Performance}\label{sec:payload}
\subsection{Overview}\label{sub:payload:overview}
\xcalibur is a scattering polarimeter in the focal spot of a focusing X-ray mirror.
X-rays are focused onto a beryllium cylinder, in which they scatter with a probability of \SI{65}{\percent} at \SI{18}{keV} and \SI{>85}{\percent} at \SIrange{30}{60}{keV}.
The scattered X-rays are then detected by CdZnTe (CZT) detectors surrounding the scattering element on four sides with an efficiency of \SI{\sim 83}{\percent} above \SI{20}{keV}, which is almost entirely determined by the geometry of the arrangement~\cite{kislat_etal_jatis_2018}.
X-rays that do not interact with the scattering element or scatter in a forward direction are detected by an imaging CZT detector facing the mirror.
The polarimeter is described in more detail in Section~\ref{sub:polarimeter}.

\begin{table*}
  \centering
  \caption{\xcalibur specifications and performance.}
  \begin{tabular}{lp{.4\linewidth}p{.35\linewidth}}
   \toprule
    Component & Description & Performance \\
   \midrule
    Truss & Carbon fiber tubes and aluminum joints & Focal spot movement \SI{<3}{mm} \\
    Pointing system & Pitch-yaw articulated & Pointing precision 1.0--\ang{;;3.6} ($3\sigma$) on source \\
    Star camera & \SI{100}{mm}, f/1.5 short-wave infrared lens & Pointing knowledge ${<}\ang{;;15}$ ($3\sigma$) \\
    X-ray mirror & Wolter I, \SI{8}{m} focal length, diameter \SI{40}{cm}, 255 Pt-C coated shells & Effective area \SI{93}{cm^2} at \SI{20}{keV} \\
    Polarimeter & Beryllium scatterer, 17 CZT detectors (each: \SI{2x20x20}{mm}, 64 pixels), NCIASIC2 readout & Bandpass: \SIrange{15}{50}{keV}; $\Delta E(\SI{40}{keV}) = \SI{6.2}{keV}$ FWHM \\
    Power & Detector / Detector + Heaters & \SI{40}{W} / \SI{160}{W} \\
    Mass & Mass under rotator / Total suspended mass & \SI{1626}{kg} / \SI{2411}{kg} \\
   \midrule
    Signal rate & 1 Crab source at \ang{45} (\ang{60}) elevation & \SI{.61}{Hz} (\SI{0.88}{Hz}) at \SIrange{15}{50}{keV} \\
    Background rate & CsI shield veto applied & \SI{5.7}{Hz} at \SIrange{15}{50}{keV} \\
    Modulation factor & Energy-independent & \num{0.51} at \SIrange{15}{50}{keV} \\
    MDP (\SI{99}{\percent} CL) & 1 Crab source at \ang{45} elevation & $11.8\% \; t_\text{day}^{-1/2}$ \\
   \bottomrule
  \end{tabular}
  \label{tab:summary}
\end{table*}

The mirror with a focal length of \SI{8}{\meter} was developed for the \infocus hard X-ray imaging telescope~\cite{tueller_etal_exa_2005}.
It consists of a pair of 255 nested Al shells with between 25 and 65 Pt-C multilayer coating pairs in order to achieve an effective area of \SI{93}{\square\centi\meter} at \SI{20}{keV} and \SI{30}{\square\centi\meter} at \SI{40}{keV}.
The mirror has a field of view of \SI{10}{arcmin} FWHM and an angular resolution of \SI{2.5}{arcmin} Half Power Diameter (HPD)~\cite{okajima_etal_apo_2002,berendse_etal_apo_2003,ogasaka_etal_2008} resulting in a diameter of the point spread function of \SI{5.8}{mm} in the focal plane.

Mirror and detector are supported by an \SI{8}{\meter}-long optical bench~(Section~\ref{sub:truss} and Ref.~\cite{kislat_etal_jai_2017}) pointed with arcsecond stability by the Wallops Arc-Second Pointer (WASP, Section~\ref{sub:wasp} and Ref.~\cite{stuchlik_2015}).
The Camera Attitude and Reference Determination System (CARDS) star tracker acting in concert with the LN251 inertial navigation system provides absolute pointing information with a precision of~\num{0.4}~arc seconds ($3 \sigma$).

\begin{figure}
    \centering
    \includegraphics[width=\linewidth]{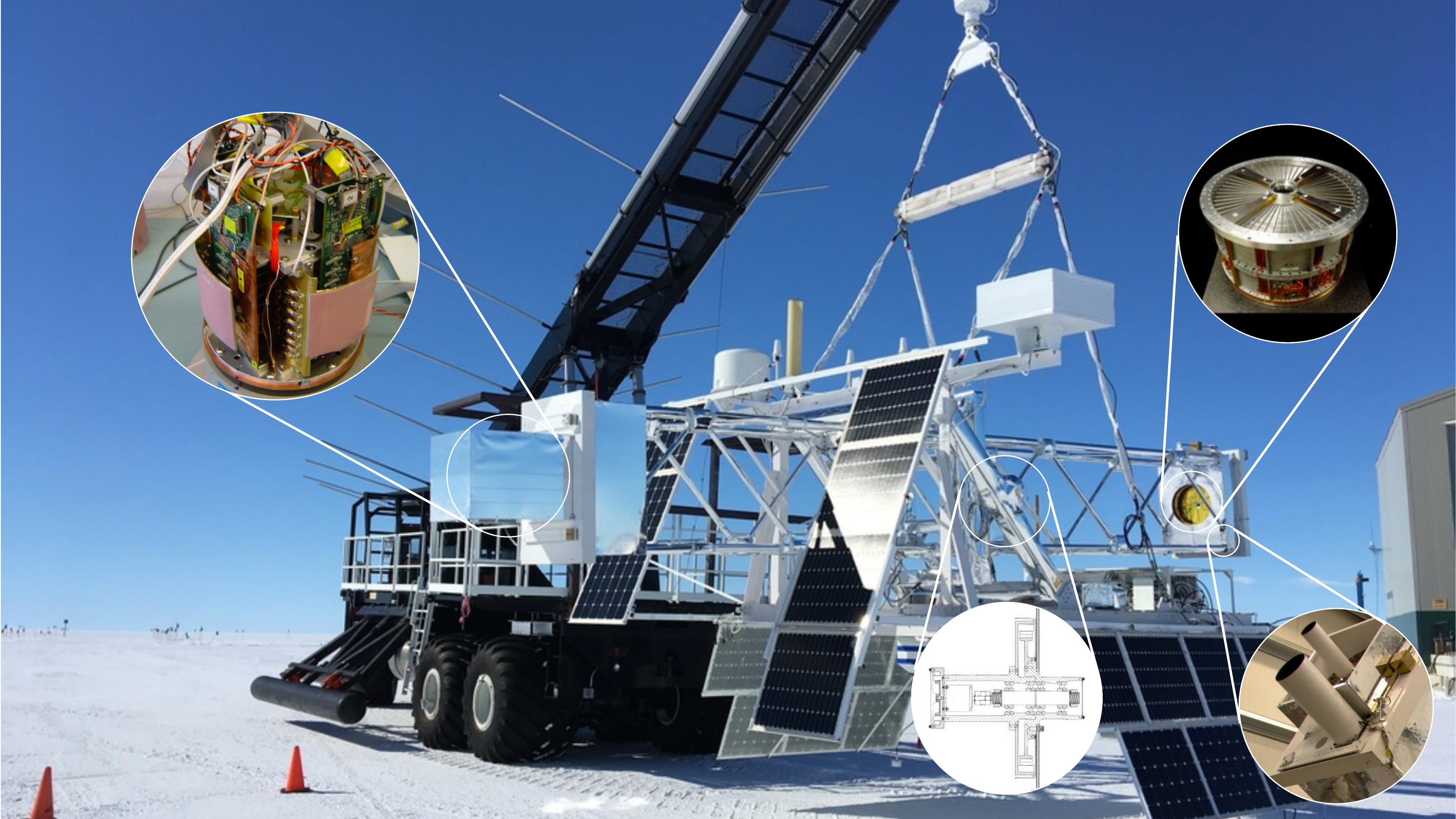}
    \caption{The \xcalibur payload and its main components during the final compatibility test at the McMurdo Long-Duration Balloon Facility. Highlighted from left to right: the \xcalibur polarimeter at the focal point of the X-ray telescope; the WASP torque motors used to point the system; the \infocus X-ray mirror; the CARDS star tracker.}
    \label{fig:telescope}
\end{figure}

Table~\ref{tab:summary} summarizes \xcalibur payload characteristics and performance.

\subsection{Polarimeter}\label{sub:polarimeter}
\begin{figure*}
  \centering
  \includegraphics[width=.85\textwidth]{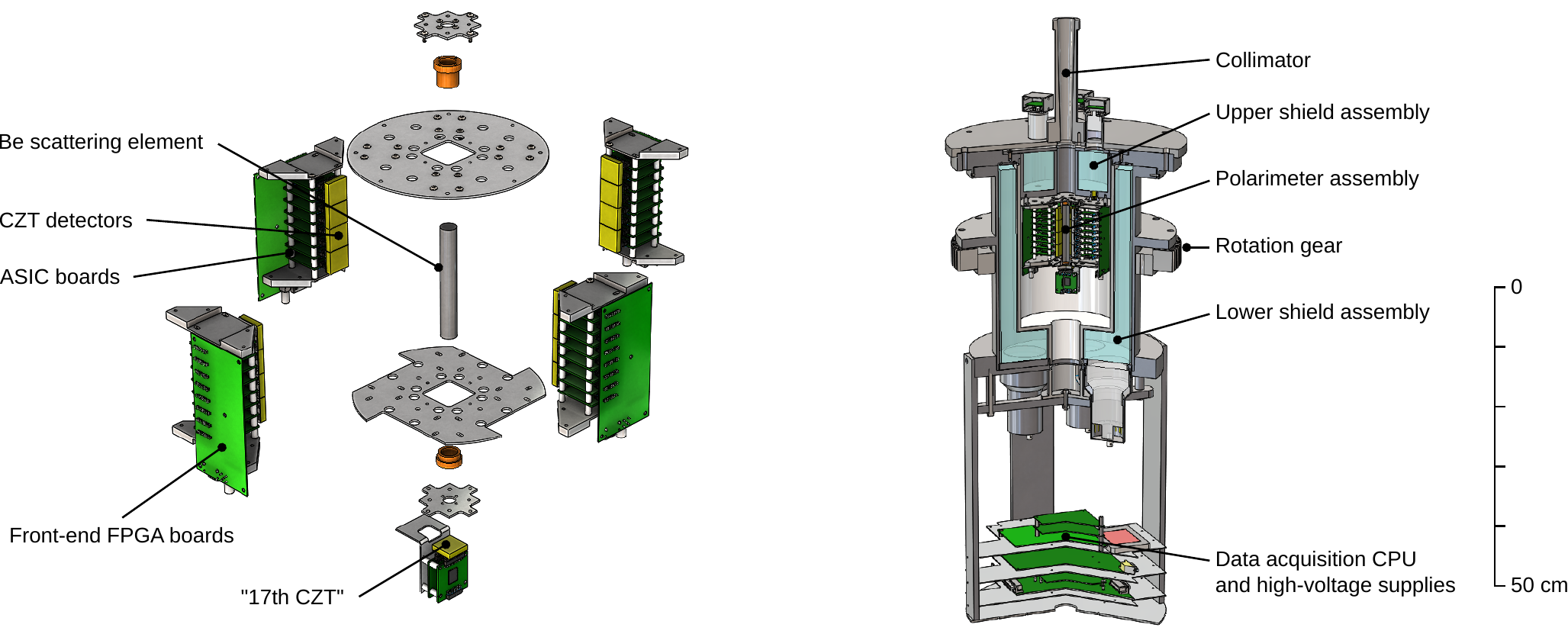}
  \caption{\textit{Left:} Exploded view of the \xcalibur detector. \textit{Right:} Complete polarimeter assembly including anti-coincidence shield and readout electronics.}
  \label{fig:polarimeter}
\end{figure*}

The polarimeter~\cite{beilicke_etal_jai_2014,kislat_etal_jatis_2018} shown in Fig.~\ref{fig:polarimeter} consists of a cylindrical, \SI{8}{cm} long beryllium scattering element with a diameter of \SI{1.2}{cm} surrounded by 16 Cadmium-Zinc-Telluride (CZT) detectors with a size of \SI{2x2x0.2}{\centi\meter}.
The Be element is aligned with the optical axis of the X-ray mirror, at the focal point.
Due to the low atomic number of Be, X-rays preferentially scatter and are then detected by the CZT detectors\footnote{An earlier version of \xcalibur used an active scintillator scattering element. However, a detailed trade study found a better senstivity when using a passive Be element~\cite{kislat_etal_jatis_2018}.}.
A \nth{17} CZT detector facing the X-ray mirror is mounted behind the scattering element to detect photons that have not scattered in order to verify the alignment of the detector with the X-ray optics.
Signals in the 64 pixels of each detector are read out by two 32-channel NCIASIC2 ASICs~(application specific integrated circuits,~\citep{wulf_etal_nima_2007,deGeronimo_etal_ieee_2008}).
The detector is surrounded by a CsI(Na) anti-coincidence detector for background rejection.
During the flight, the polarimeter assembly rotated at 3~rpm in order to reduce systematic errors due to azimuthal asymmetries of the detector performance.
The performance of the polarimeter will be described in more detail in Section~\ref{sec:performance}.

The anti-coincidence shield consists of two parts~(Fig.~\ref{fig:polarimeter}, \emph{right}).
The bottom part is bucket-shaped with a CsI thickness of \SI{27}{mm} on the side and \SI{50}{mm} on the rear.
This part of the CsI shield is read out by four Hamamatsu R6233-100 PMTs with a diameter of \SI{3}{in}.
The cable opening is covered by a small, \SI{6.4}{mm} thick tungsten shield.
The top part of the shield is made of \SI{50}{mm} thick CsI and is additionally covered by an \SI{8}{mm} thick tungsten plate with an attached tungsten collimator around the X-ray entrance window.
The top part of the active shield is read out by four Hamamatsu R1924A, \SI{1}{in} diameter PMTs.
Signals in the four PMTs of each shield part are summed and threshold discriminators generate a veto signal for each of the shield parts.
This signal is extended digitally to a programmable duration set to \SI{6.4}{\micro\second} based on a study of correlations between PMT and CZT signals.
The detector volume was light-tightened, with a \SI{12.7}{\micro\meter} thick aluminum foil covering the entrance window, because we found that sunlight can increase the noise in the CZT detectors due to increased leakage current.
The performance of the shield is described in Section~\ref{sub:shield}.

The CZT detectors are read out by a total of \num{34} ASICs.
Each channel of an NCIASIC2 consists of a charge-sensitive pre-amplifier, operated during the \xcalibur flight at a nominal gain of \SI{57}{\milli\volt/\femto\coulomb}, a 5th order shaping amplifier operated with a time constant of \SI{1}{\micro\second}, a peak amplitude detector with analog memory, and a threshold discriminator.
The circuit boards carrying individual ASICs are plugged into four front-end boards with FPGAs that control the readout and digitization of signals from the ASICs.
These boards are connected to a PC/104 computer outside the anti-coincidence shield using low-voltage differential signalling (LVDS).
In order to minimize noise pickup, all digital signals are suspended during data acquisition.
After a trigger in any ASIC, data acquisition remains active for \SI{2}{\micro\second}, after which all ASICs are disabled, and the readout clock is enabled after an additional \SI{5}{\micro\second} delay.
The peak voltage in up to three triggered channels per ASIC is digitized with 12-bit precision and the result is transmitted to the PC/104 computer.
Data acquisition is re-enabled~\SI{17.2}{\micro\second} after completion of the readout.
This delay is necessary because we found unacceptable noise picked up by the ASIC front-end when enabling acquisition immediately after the last clock cycle.
Altogether, readout of a single-pixel event results in a detector dead time of \SI{43.6}{\micro\second}, with an additional \SI{20}{\micro\second} per pixel within the same ASIC.
At most 3 channels are read out per ASIC, and the longest possible dead time per ASIC is \SI{83.8}{\micro\second}.
The four front-end boards read out ASICs and transmit data in parallel.

Reading data from the PC/104 into memory on the data acquisition computer takes roughly \SIrange{40}{60}{\micro\second} per event.
On the PC/104 board, one FIFO buffer per front-end board allows buffering data from up to two ASICs, which can be part of the same event or from separate events.
Absolute timing is provided by the WASP GPS receiver.
A pulse-per-second signal from this receiver resets a local \SI{1.28}{\micro\second} counter on the PC/104 board every second and a second counter counts these pulses.
Both counters are latched into registers for subsequent software readout every time data from an ASIC are received by the PC/104 board.
The two parts of the anti-coincidence shield each produce a veto flag with a duration of \SI{6}{\micro\second}, whose values are stored in the input buffer when data from an ASIC are received.
Finally, an optical encoder registers the rotation angle of the polarimeter with \ang{1} precision.
Further details of the data acquisition, and the dead time during flight are discussed in Section~\ref{sub:deadtime}.

In order to reduce dead time and data rate, we implemented a ``sparse'' mode.
In this mode, readout and digitization are cancelled if an event is vetoed by a signal in the anti-coincidence shield.
This reduces the dead time to \SI{\sim 27}{\micro\second} per event.
During most of the flight, the full data were archived and the sparse mode was not used.

\subsection{Truss}\label{sub:truss}
The \infocus X-ray optics and the polarimeter are supported by an \SI{8}{m} long optical bench~\citep{kislat_etal_jai_2017} (typically referred to as telescope truss).
This truss consists of two sections made of carbon-fiber tubes glued to machined aluminum joints.
The two sections are bolted to a welded aluminum center frame.
Aluminum honeycomb panels holding equipment are bolted to the ends of the truss.
The entire truss structure is painted white and wrapped in Dacron mesh and aluminized Mylar for thermal control.

The goal of the design is to minimize thermal and mechanical deformations,
ensuring that the focal point of the X-ray mirror remains 
close to the center of the Be scattering slab.
Measurements and calculations have shown that the systematic error on the polarization fraction will be~\SI{<1}{\percent} if the focal spot is within~\SI{1}{mm} of the center of the scattering element.
If deflections are~\SI{<3}{mm}, corrections can be applied to reduce the systematic error to~\SI{<1}{\percent}~\citep{beilicke_etal_jai_2014}.
During the flight, we monitored the location of the focal spot with a camera mounted in the center bore hole of the mirror observing a pattern of LEDs mounted to the detector~\citep{kislat_etal_jai_2017}.
An image is captured automatically every~\SI{20}{s} and analyzed on the payload in order to extract the location of the focal point relative to the detector.
This information is transmitted to the ground as part of the regular housekeeping data.
We also have the ability to transmit complete images for diagnostic purposes.

\begin{figure}
  \centering
  \includegraphics[width=\linewidth]{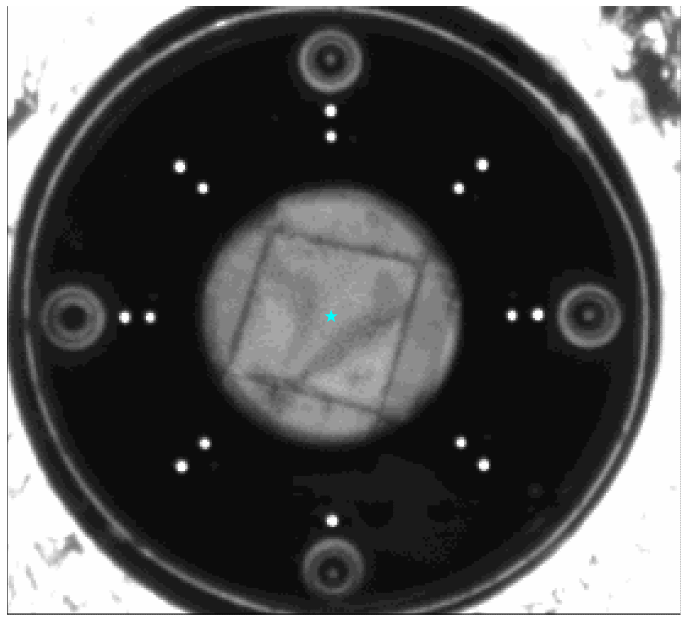}
  \caption{Sample image from the \xcalibur alignment monitoring camera. Images with \num{1360x1024} gray-scale pixels are being recorded and analyzed on the payload, but images are cropped to the region containing the LED pattern for storage and transmission. 
  The 15 round bright spots correspond to the LEDs.
  The single LED at the bottom of the pattern is used as an index to measure the rotation angle of the pattern.
  The cyan star indicates the reconstructed center of the LED pattern.
  Additional information about the detected clusters and fit results are transmitted, but are not shown for clarity.
  All LEDs were detected at their correct locations.}
  \label{fig:alignment_sample}
\end{figure}

Figure~\ref{fig:alignment_sample} shows an example of an image taken during flight.
The camera is equipped with a \SI{100}{mm} lens and records images with a resolution of \num{1360x1024} pixels.
For transmission and storage, images are cropped to a region of interest around the LED pattern, but the image analysis considers the entire image.
The LEDs are mounted to a disk-shaped circuit board painted black and shaded from the sun with the help of a baffle.
The image analysis begins by finding this circuit board as an annulus of dark pixels on the bright background.
The bright region at the center of the image is the entrance window to the polarimeter, covered in thin aluminum foil and kapton tape.
Within the annulus, LEDs are detected as clusters of bright pixels.
The known pattern of LEDs is then fitted to the distribution of these clusters by shifting, rotating, and scaling the pattern in order to minimize the distances between LED positions in the pattern and clusters in the image.

\begin{figure}
  \centering
  \includegraphics[width=\linewidth]{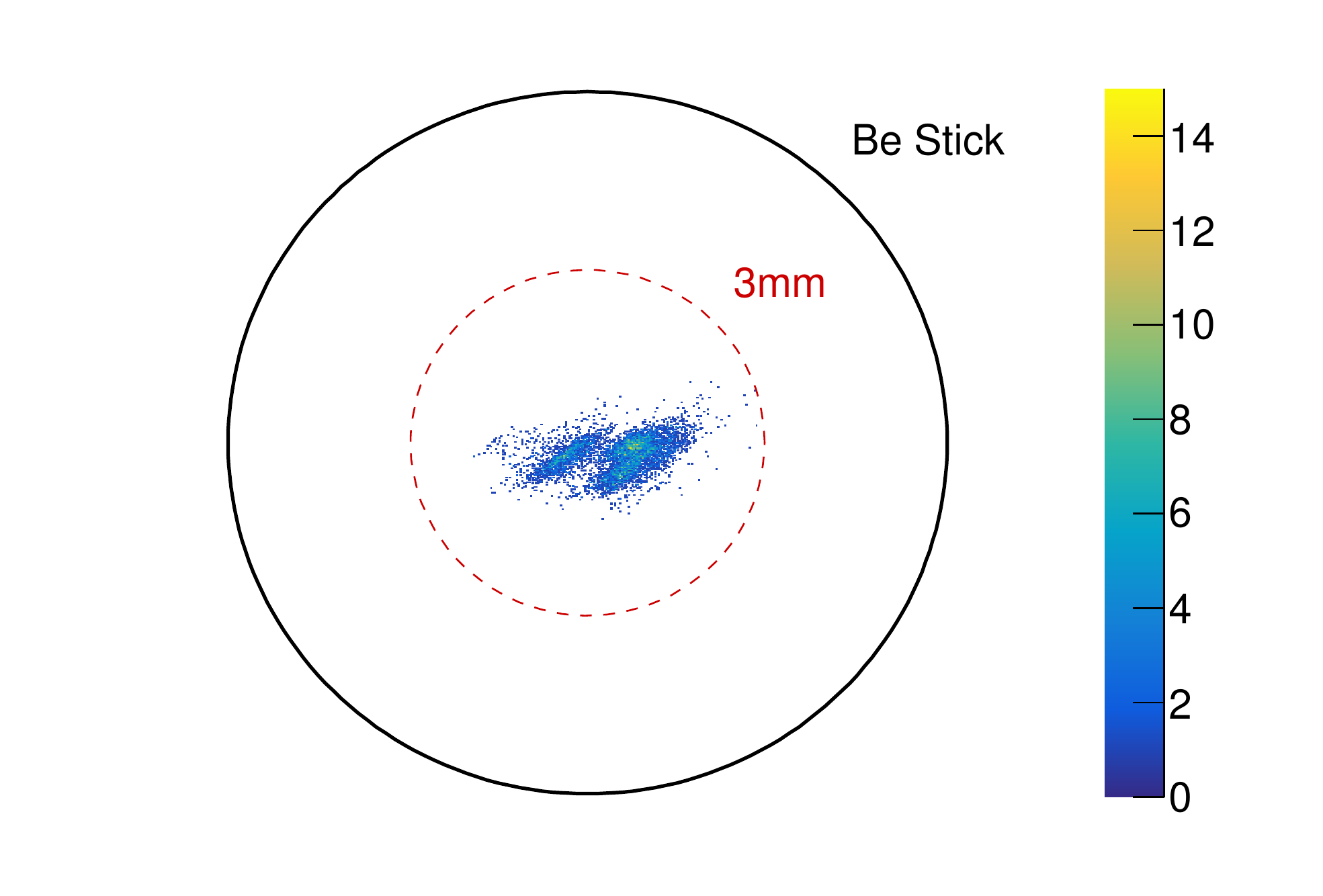}
  \caption{Measured locations of the focal spot relative to the Be scattering element. The black circle indicates the diameter of the scattering element, the red dashed circle has a radius of \SI{3}{mm} around the center of the scattering element, and the histogram color indicates the number of alignment measurements.}
  \label{fig:alignment_circle}
\end{figure}

Through the entirety of the flight, we found an RMS spread of the focal spot location of~\SI{0.63}{mm} horizontally and~\SI{0.28}{mm} vertically~(see Fig.~\ref{fig:alignment_circle}).
All alignment measurements were~\SI{<3}{mm} from the center of the scattering element, and the focal point was within~\SI{1}{mm} of the center~\SI{73}{\percent} of the time.
We also observe a significant correlation between horizontal alignment offset and pointing elevation, and the horizontal RMS spread during the GX 301-2 and Vela X-1 observations is~\SI{0.47}{mm} and~\SI{0.30}{mm}, respectively.
The fact that changes in pointing elevation result in a horizontal instead of vertical offset has been observed in pre-flight tests of the truss structure and we attribute it to torsional deformations.
Since the truss structure has been cut during the recovery of the polarimeter after the flight, we do not have the ability to verify this hypothesis.

\begin{figure}
  \centering
  \includegraphics[width=\linewidth]{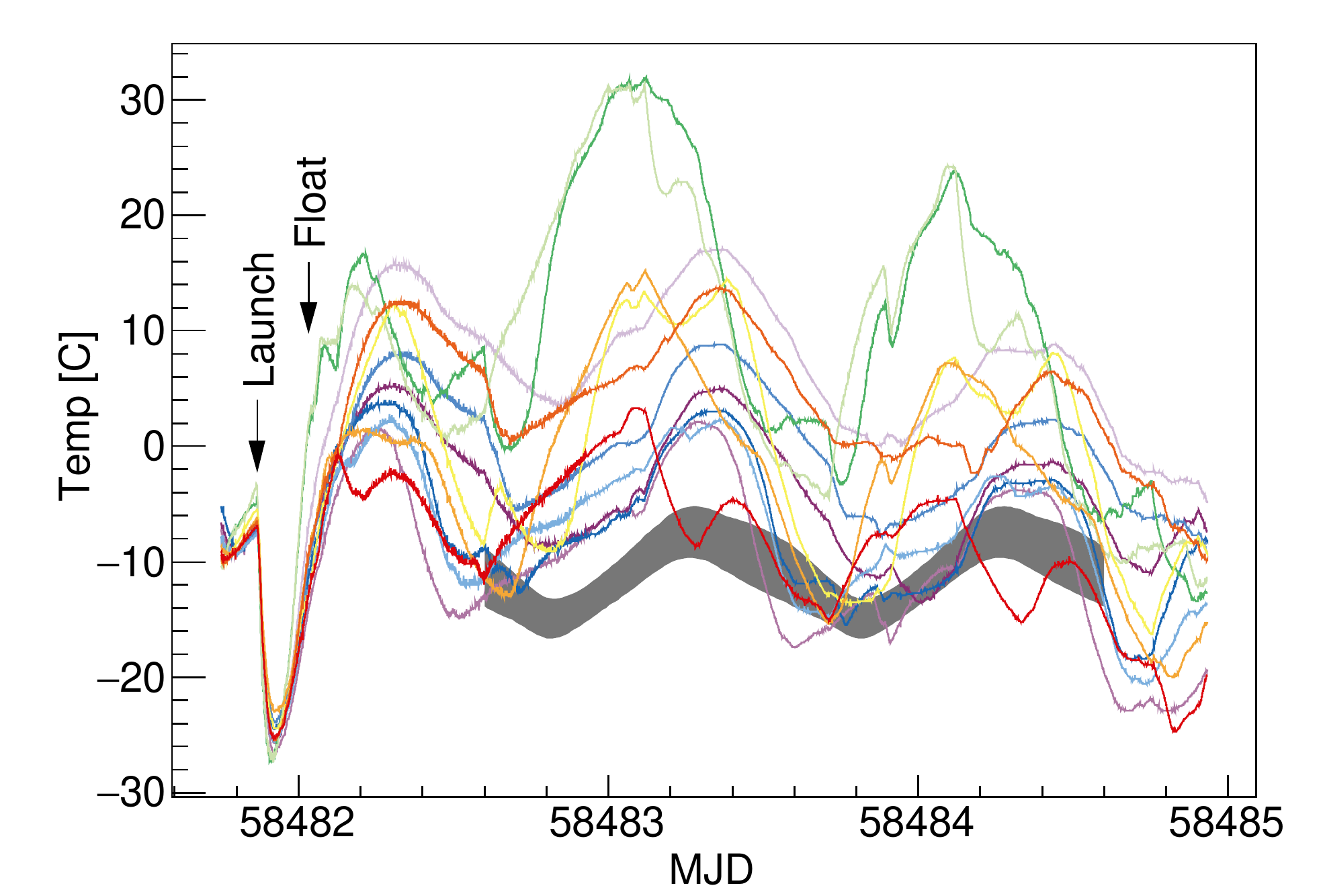}
  \caption{Temperature of truss joints measured during the flight. Each curves represents one of the 12 aluminum joints. The grey shaded area indicates the range of predictions for the simulated ``warm'' case during an observation of GX301-2. The two joints that hold the latch bar (green curves) show greater variation because the latch bar was not wrapped in aluminized mylar. See Section~\ref{sub:truss} for details.}
  \label{fig:joint-temperatures}
\end{figure}

Figure~\ref{fig:joint-temperatures} shows the temperature of each of the~12 aluminum joints throughout the flight.
After the launch, joint temperatures reflect the decreasing air temperature until the balloon crosses the tropopause.
Temperatures start diverging after the balloon reached the floating altitude of \SI{130000}{ft} due to the low air density.
At this altitude temperature is determined by solar illumination and exposure of each joint to the ground or cold cloud cover.
The two joints that hold the latch bar show greater variation because the latch bar was not wrapped in aluminized mylar as safety precaution.
Mylar could interfere with the latch, if were to come lose during launch.
The latch bar holds the mechanical assembly that arrests the truss in its stowed position during ascent and decent.

During the design phase the anticipated truss configuration was analyzed in Thermal Desktop, RadCAD, and SINDA~\cite{kislat_etal_jai_2017}.
Two cases were analyzed: a hot case with an environment representative of the conditions near the December solstice and a cold case representing the conditions at the end of January.
Since the entire flight took place close to the solstice, only the hot case is shown in Fig.~\ref{fig:joint-temperatures}.
Disregarding the extremes due to the exposed latch bar, which was not modelled, and data towards the end of the flight when the balloon was at a very low altitude, three main observations can be made: (1) most joint temperatures were higher than predicted; (2) the large spread of temperatures between individual joints of about \SI{20}{\degreeCelsius} between the warmest and coldest joints was not anticipated; (3) the model underestimates the observed diurnal variation. 
These facts indicate that the performance of the insulating layers on the truss was overestimated in the model.
However, despite this, the telescope achieved the required alignment performance.

\subsection{Pointing System and Pointing Performance}\label{sub:wasp}
The Wallops Arc Second Pointer (WASP) was utilized to provide attitude control for the 8-m truss-mounted instrument. 
The WASP is a balloon-borne fine-pointing system designed to point telescopes and other instruments on balloon gondolas
at targets with arc-second accuracy and stability~\citep{stuchlik_2015}. The WASP system points an instrument using a 
gondola-mounted pitch/yaw articulated gimbal. The range of motion of the yaw-gimbal is purposely minimized to reduce 
kinematic coupling during fine pointing. Thus, the gondola itself is suspended beneath a standard NASA Rotator to provide 
large angle azimuth targeting and coarse azimuth stabilization.  

Instrument attitude is computed by integrating incremental angles (delta-Theta) output from an LN251 system. Quaternion solutions
from a Camera Attitude and Reference Determination System (CARDS) star tracker are augmented by the integrated attitude solution utilizing
a 6-state Extended Kalman Filter. The CARDS was developed as a WASP subsystem to be robust in the face of significant bright-sky  background, typical of daytime balloon flight operations. 

This flight was the \nth{7} flight of the WASP system and was the second flight
with the \xcalibur payload. However, the Antarctica flight provided a number of 
firsts for WASP. It was the first multi-day flight of WASP. It was first the time WASP 
performed an in-flight alignment between the star tracker, inertial measurement
unit, and a science instrument. It was the first flight of WASP integrated with 
the improved rotator electronics system. It was the first attempt at fully autonomous
control system operations. 

The in-flight alignment of the star tracker with the instrument was achieved by comparing the star tracker image to a forward-looking camera mounted to the central borehole of the X-ray mirror.
The alignment between this on-axis camera and the X-ray mirror had been calibrated in the lab.

The primary science requirement levied on the WASP was to hold the X-ray mirror axis to within 30 arc seconds (3$\sigma$) of the
source during each observation. 30 arc seconds 
correspond to 1.16~mm in the focal plane of the telescope.
During the flight, WASP maneuvered to and acquired multiple X-ray sources and provided pointing stability 
between 1.0 and 3.6 arc seconds (3$\sigma$) while on the source. 

\begin{figure}
    \centering
    \includegraphics[width=.5\textwidth]{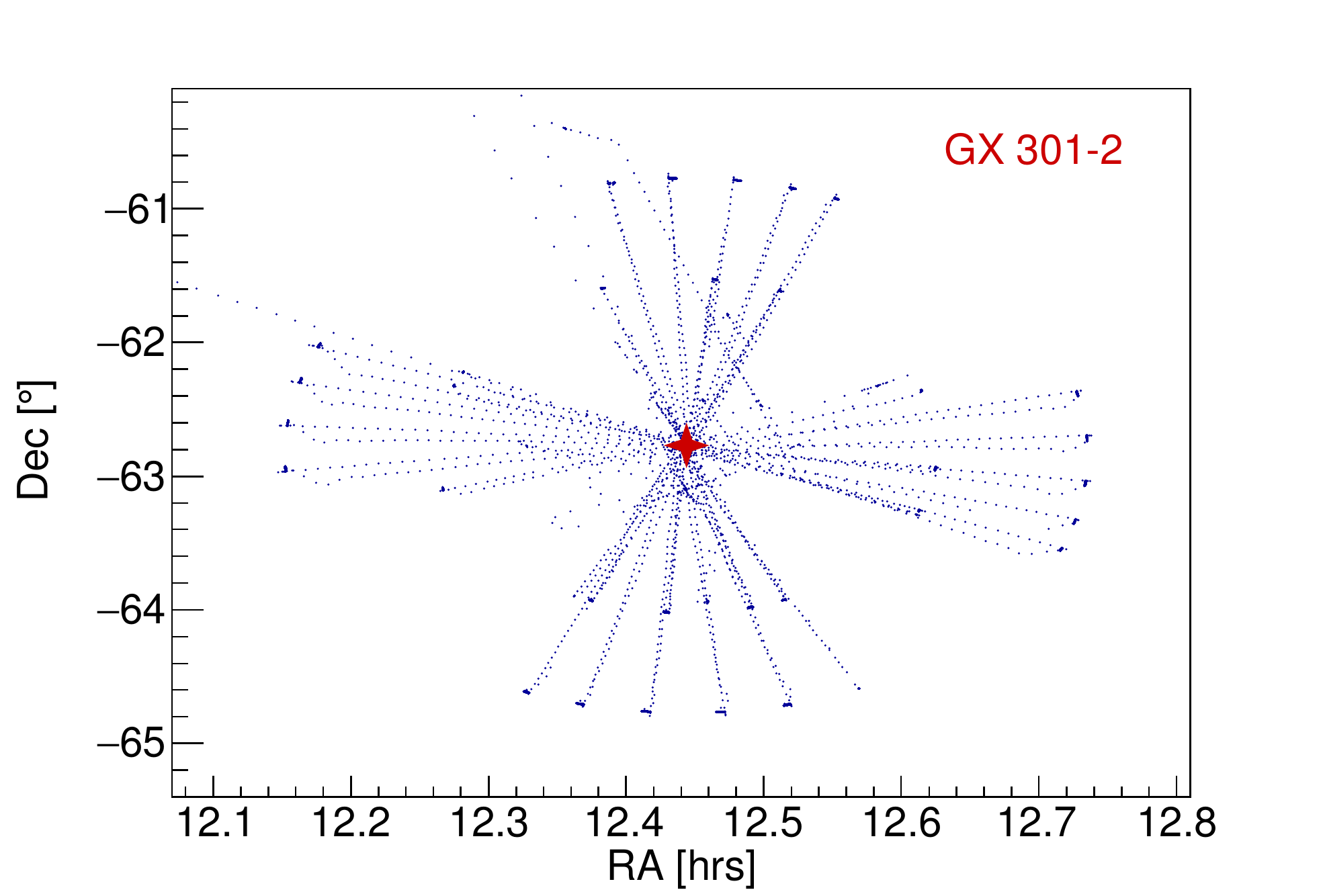}
    \caption{Pointing pattern during an observation of GX 301-2. Cycling between on and off source observations with a period of \SI{15}{min} allows measurement of the changing background.
    During the observation the magnitude of off-source slews was reduced from \ang{2} to \ang{1}, which is still significantly more separation than the field of view of the X-ray optics.}
    \label{fig:pointing_pattern}
\end{figure}

WASP pointing was spent primarily on two targets: GX 301-2 and Vela X-1. The targets
were complementary in that there were no times when both sources violated a lower or upper
look elevation constraint. There was always one target available for observation. Pointing at each target followed a ``nodding'' program
where the control system cycled the instrument between time on source and time on background at varying locations 
relative to the source (pitch up/down, yaw right/left, see Fig.~\ref{fig:pointing_pattern}).
The frequency of this nodding pattern was adjusted to the observed sources.
For the pulsars Vela~X-1 and GX~301-2, it was chosen such that each on-source observation would be longer than one pulsar period (\SI{\sim10}{min} and \SI{\sim15}{min}, respectively).
Frequent off-source observations are necessary to measure the background, which may change with pointing elevation, balloon altitude and time.

\begin{figure}
  \centering
  \includegraphics[width=\linewidth]{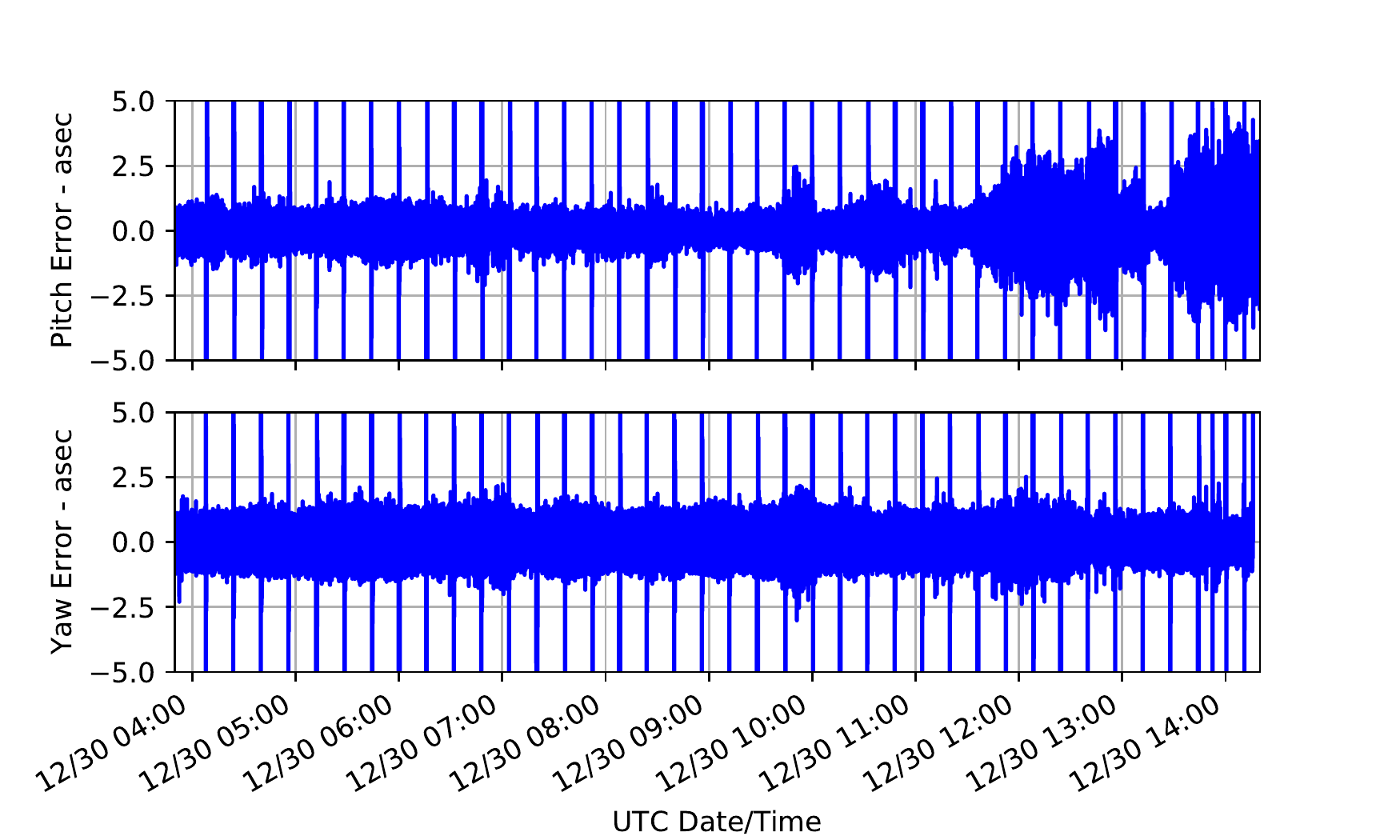} \\
  \includegraphics[width=\linewidth]{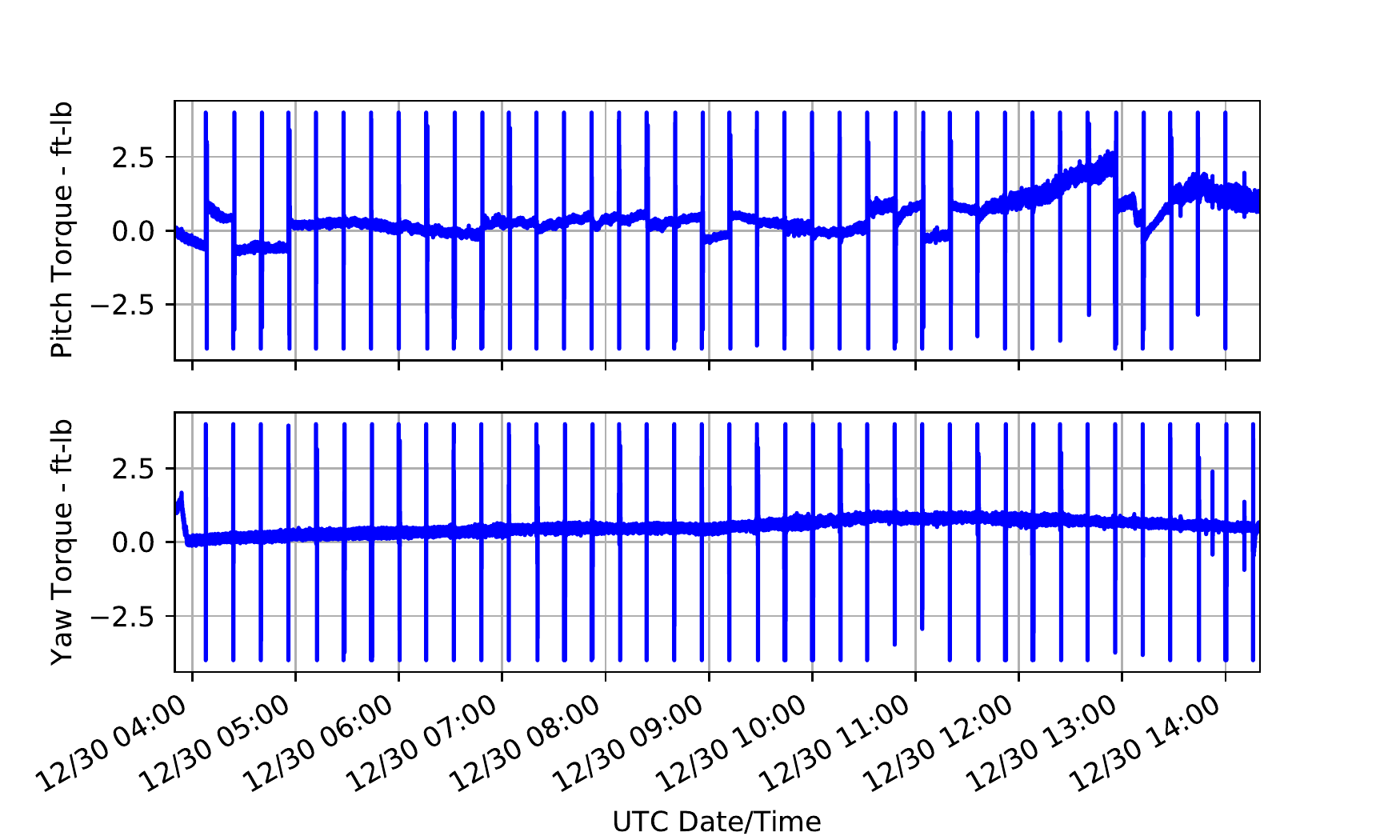}
  \caption{
    \emph{Top:} Raw controller errors during \nth{1} observation of GX301-2.
    \emph{Bottom:} Controller response torques during \nth{1} observation of GX301-2.
    In both figures, the large excursions correspond to slews between on-source and off-source observations (see Fig.~\ref{fig:pointing_pattern}).
  }
  \label{fig:gxrawpointing}
\end{figure}

Figure \ref{fig:gxrawpointing} (top) shows the pointing error experienced by the control system during the first observation of
GX 301-2 as the system cycled on and off of the source. The range of pointing performance experienced during this observation 
was typical for that of the flight. While on source, the yaw error was generally kept within 1.5 arc seconds. 
The pitch error performance varied between 1 and 1.5 arc seconds for most of the observation.
At the higher elevations experienced toward the end of the observation, pitch error and pitch torques increased (see
Fig.~\ref{fig:gxrawpointing}, bottom).
We interpret this as an indication, that the system picked up some mechanical interference between one of the wire cable 
loops and a gimbal.
Even still, pointing stability was maintained well within the experiment requirements.

\subsection{Mirror -- detector alignment}
To verify in the field that the mirror is aligned with the polarimeter, we developed a portable system that attaches to the front of the telescope. 
It consists of a \SI{14}{inch}-diameter Celestron off-the-shelf Schmidt-Cassegrain telescope operated as a light collimator rather than collector: at the focal point, we position the end of an optical fiber pigtailed to a Thorlabs \SI{637.4}{nm} Class 3R laser.

\begin{figure}
    \centering
    \includegraphics[width=0.8\linewidth]{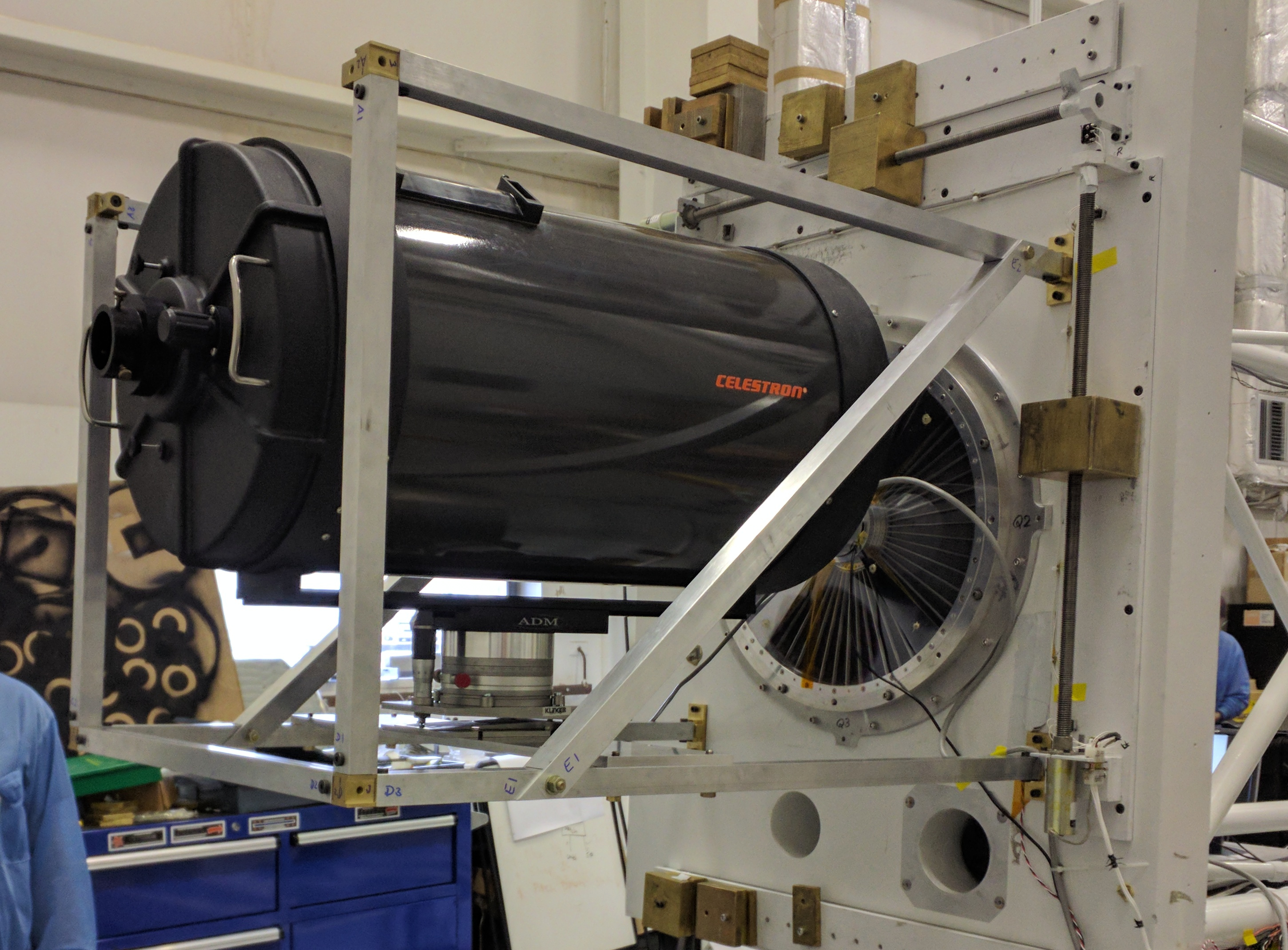}
    \caption{The in-field aligner mounted to the front bulkhead in front of the X-ray mirror. The tilt and rotation stages are seen between the telescope and the mounting structure, and the eyepiece hole where the laser fiber goes (not in this image) is on the left. The telescope is slightly offset vertically from the center of the X-ray mirror so that the forward looking camera sees the collimated beam rather than the secondary Schmidt-Cassegrain mirror.}
    \label{fig:aligner}
\end{figure}

The laser is bright enough that the focal spot can be observed without darkening the room, and the telescope is large enough that when mounted in front of the X-ray mirror it illuminates a significant fraction of the mirror.
As the optical telescope does not illuminate the entire X-ray mirror, 
the image of the light source in the detector plane does not 
fully correspond to the image of a celestial source, and the position
may be offset slightly. The offset will be smaller than the angular 
resolution of the mirror.
We coarsely collimate the light from the telescope by adjusting the telescope focus so that at some large distance the beam is still \SI{\sim14}{inches} in diameter. 
Then, after mounting the telescope to the front bulkhead, we finely adjust the focus to optimize the point spread function seen by the forward looking camera. 
The telescope is mounted on top of tilt and rotation stages to adjust the position of the light source in the forward looking camera to match its position during calibration in the X-ray optics lab at Goddard Space Flight Center.
By then comparing the position of the focal spot on the detector in the back-looking camera to its calibration, we can verify that the X-ray mirror is aligned with the detector.
Using this technique, we verified that the X-ray mirror was aligned within \SI{0.39}{mm} of the center of the Be scattering element.

\begin{figure}
    \centering
    \includegraphics[width=0.8\linewidth]{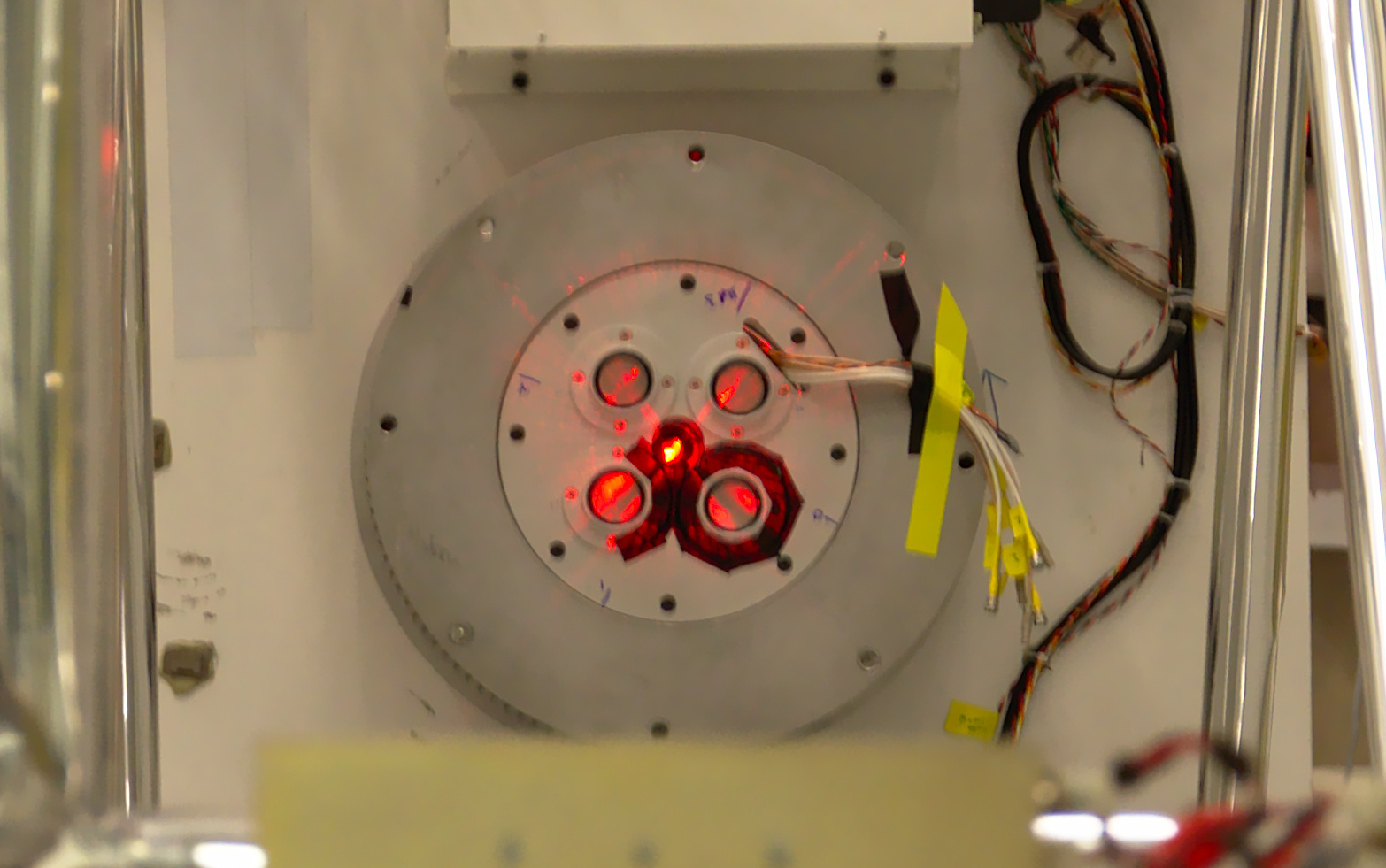}
    \caption{The X-ray mirror's focal spot on the detector, as illuminated by the in-field aligner. The photograph shows the shield assembly enclosing the polarimeter from above with the collimator removed (see Fig.~\ref{fig:polarimeter}).}
    \label{fig:focalspot}
\end{figure}

In order to allow projection of measured polarization angles onto the sky, the roll angle of the polarimeter was calibrated with respect to the pointing system's bank angle as follows. With the truss oriented in a stow position (with a truss elevation at 6$^{\circ}$), the polarimeter was rotated such that one of the
detector segments was oriented approximately vertically. 
Two angles were measured simultaneously:  the offset of the pointing system's 
roll angle from the direction to the center of the earth (reading out the LN 251), and the offset of the polarimeter's side from the direction of to the center of the earth (measured with the help of an engineer's spirit level 
and the back looking camera taking images of the detector and the level).
These two measurements allowed us to reference the polarimeter's roll angle
relative to the pointing system's roll angle. During the flight, the pointing system then reports the roll angle offset to the celestial north pole, and the polarimeter reports its roll angle. Together with the ground calibration results, the angle between the celestial north and the detector 
orientation can be determined. For each detected event, we can thus infer the 
angle $\chi$ between the scattering direction and the celestial north pole.

\subsection{Telemetry}\label{sub:telemetry}
The data acquisition CPU is part of the continuously rotating polarimeter assembly.
A second CPU mounted on the pointed optical bench communicates with the data acquisition CPU via a custom four-wire protocol with a bitrate of \SI{6.25}{Mbit/s} through an 8-pin rotating connector that also carries power and the pulse-per-second signal from the GPS for event time tagging.
Finally, a third CPU connected to the truss-mounted CPU via Ethernet is responsible for telemetry to the ground.
\xcalibur communicated via several channels provided provided by Columbia Scientific Balloon Facility's Support Instrumentation Package (SIP), as well as a line-of-sight downlink via a \SI{150}{kbit} digital L-band transmitter while the payload was within about \num{300} miles of the launch site.
Over-the-horizon downlink was available via two Tracking and Data Relay Satellite System (TDRSS) antennas with varying bandwidth between \SI{3}{kbit/s} and \SI{83}{kbit/s}, respectively, an IP-based Iridium Pilot system\footnote{Allocation for detector operations and polarimeter data downlink. Additional bandwidth on the Iridium Pilot system was used by the WASP pointing system.} at \SI{40}{kbit/s}, and low-rate housekeeping data via Iridium short-burst data (SBD) and dial-up if needed.
Data downlinks were used in a redundant fashion, and all except the line-of-sight transmitter were shared between detector operations and the WASP.
Configurable data prioritization and filtering was used to adapt to changing available bandwidth during the flight.
Command uplink was provided by a line-of-sight antenna from McMurdo, as well as via Iridium Pilot, short-burst data, and dial-up.
A TDRSS uplink was established once a day for higher-rate commanding, in particular with the intent to upload new observation schedules to WASP.
After landing, the payload still functioned, solar panels still provided power, and due to the unexpectedly short flight, we were able to download all data directly from the truss-mounted CPU via Iridium Pilot, recovering all data that may have been lost in downlink.
All data presented in this paper were obtained in this way.
Note, however, that we still suffered some data loss due to the connection between the rotating readout CPU and the truss-mounted CPU as discussed in Section~\ref{sub:deadtime}.

\section{Polarimeter and Shield Performance}\label{sec:performance}
\subsection{ASIC and CZT selection}\label{sub:selection}
In order to achieve the best possible detector performance, we characterized \num{83} ASICs and 19 CZT detectors available at Washington University in St.~Louis, from which we assembled the 17 CZT/ASIC assemblies used in \xcalibur.
We started by evaluating each ASIC without connecting a CZT detector.
Using an oscilloscope connected to the analog output pin of the ASIC and the ASIC's internal \SI{200}{fF} test pulse capacitor, we measured the gain of each channel, operating at the nominal setting of \SI{57}{mV/fC}, and the baseline RMS noise voltage.
Using the measured gain, we converted the RMS noise into equivalent noise charge~(ENC).
Typical ENC for a good channel is \SI{<200}{e^-}, with the best channels reaching \SI{~150}{e^-}.
We also tested the linearity of the front end and recorded a sample of waveforms, which we inspected visually.
CZT detector crystals were generally found to be of good quality, and their selection had little impact on detector performance.

Based on these results, we assembled units of CZT detectors with two ASICs each using the ASICs with the lowest noise and largest number of working channels, that did not exhibit any strong nonlinearities or unusual pulse shapes.
Each CZT detector is bonded to a ceramic circuit board using a flip-chip process and the circuit board is plugged into the ASIC boards using two 32-pin connectors.
These packages are then plugged into the digital front-end boards using another set of connectors on the ASIC board.
This modular structure allows replacement of individual ASIC boards and CZT detectors even after assembly.
All of these CZT/ASIC assemblies were then tested individually by obtaining spectra with a ${}^{152}\text{Eu}$ source to verify their performance, and in order to test the CZT detectors themselves.
Key criteria were the number of working channels, energy threshold, and energy resolution of the assembly.
Poorly contacted pixels can lead to additional inoperable channels, and leakage currents can result in increased noise, which often required deactivation of additional channels.
A small number of CZTs were found to suffer from these problems and were replaced.
Thus, we systematically assembled the 17 best-performing packages, which we subsequently used when assembling the four front-end boards.

\subsection{Calibration}\label{sub:calibration}
The \xcalibur CZT detectors were calibrated at Washington University in St.~Louis using an \SI{0.2}{\text{$\mu$}Ci}  $^{152}\text{Eu}$ calibration source.
The isotope with a half life of about 13~years emits several gamma-ray lines in the energy range of interest: a line complex consisting of the $^{152}\text{Sm}_\text{K\textalpha}$ and $^{152}\text{Sm}_\text{K\textbeta}$ X-ray lines around \SIlist{40;45}{keV} and a \textgamma-ray line at \SI{121.782}{keV}~\cite{TabRad_v2}.
The count rate due to the source exceeded background rates by about three orders of magnitude.

\begin{figure}
  \centering
  \includegraphics[width=.5\textwidth]{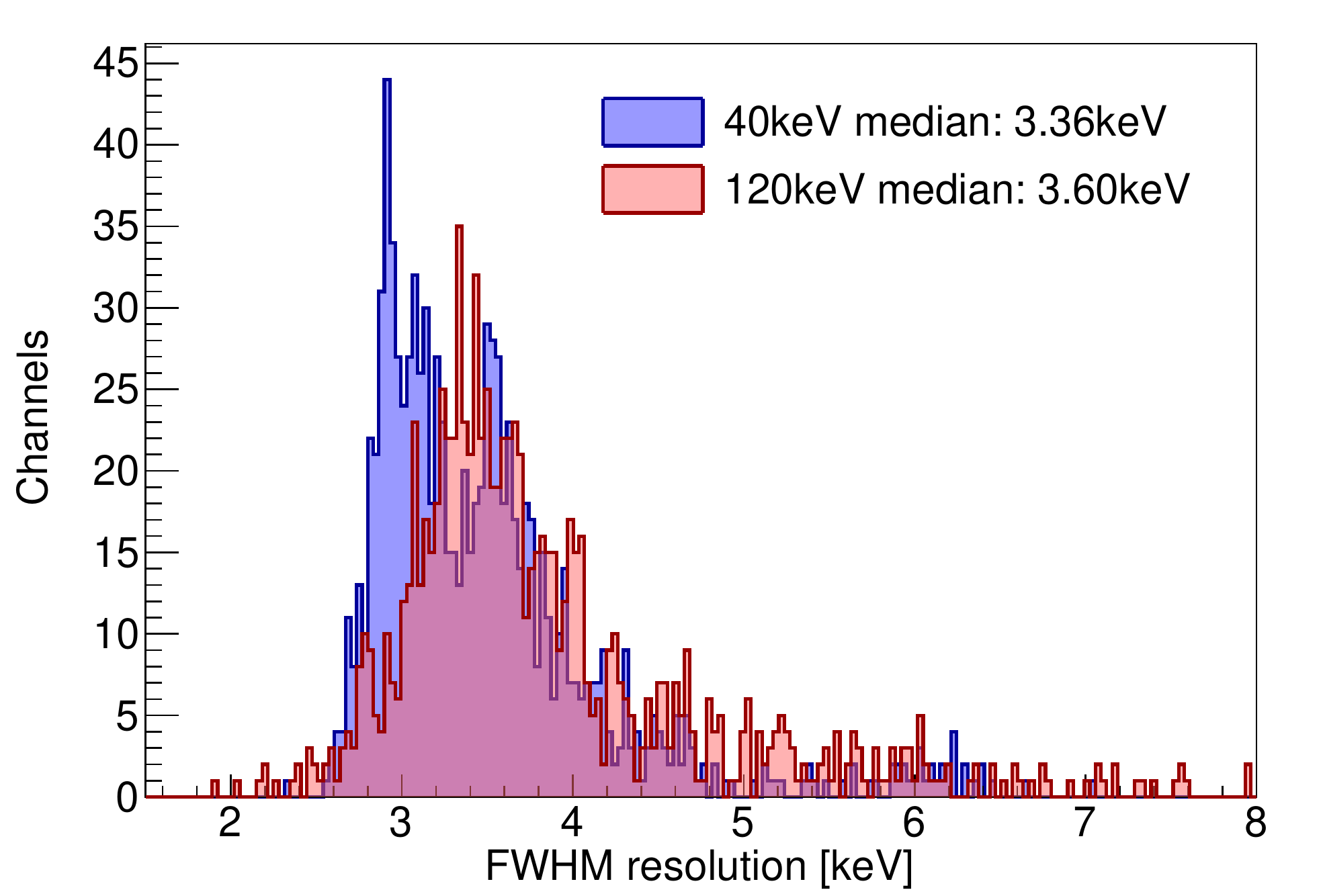}
  \caption{FWHM energy resolution near the \SI{40}{keV} K{\textalpha} line group and at the \SI{121}{keV} gamma-ray line. In general, we find a very similar resolution at both energies. The slight broadening of the distribution at \SI{121}{keV} is due to the fact that the line is generally less pronounced resulting in a larger uncertainty of the fit result.}
  \label{fig:resolutions}
\end{figure}

\begin{figure}
    \centering
    \includegraphics[width=\linewidth]{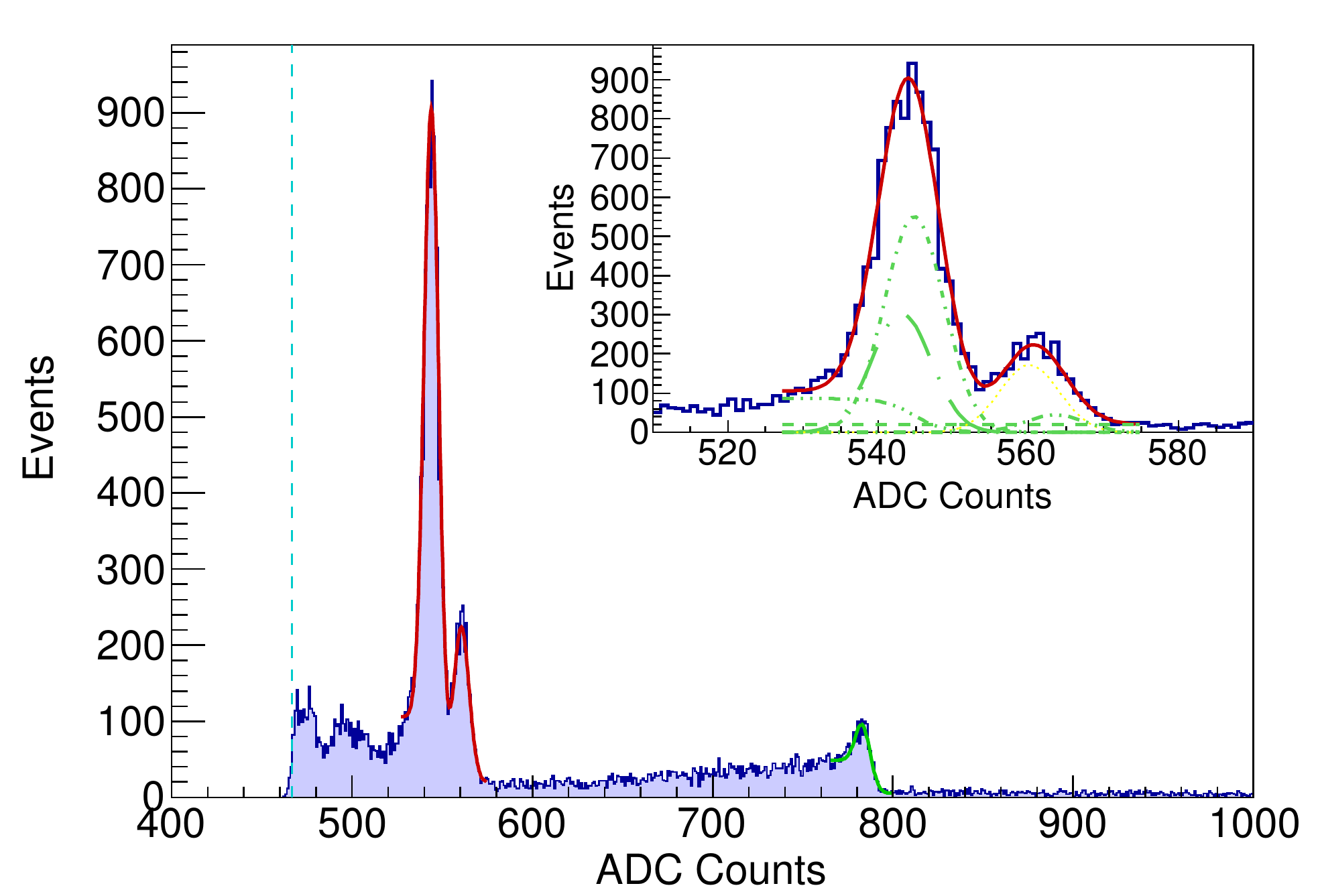}
    \caption{A representative sample calibration spectrum acquired with a single channel using a $^{152}\text{Eu}$ calibration source. The inset shows the individual components of the fit of the K-line complex between \SIrange[range-phrase={ and }]{39}{46}{keV}. The structures at energies below the K\textalpha\ line are due to escape peaks. The vertical dashed light-blue line indicates the energy threshold defined as the \num{0.25}\textsuperscript{th} percentile of the full spectrum.}
    \label{fig:spectrum28_0-detail}
\end{figure}

The \SIrange{39}{46}{keV} line complex and the \SI{121}{keV} line were jointly fit by the following two functions:
\begin{align}%
  \label{eq:f40}%
  \begin{split}%
    f_{40}(E) &= C_{40} +\frac{y_{40}}{1 + \exp\left(\frac{E-E_1}{\sigma_{40}/2}\right)} \\
      &\qquad+ A_{40}\sum_{i = 1}^3 a_i\exp\left(\frac{(E-E_i)^2}{2\sigma_{40}^2}\right)
  \end{split}\\
\intertext{and}
  \begin{split}
    f_{121}(E) &= C_{121} + A_{121}\exp\left(\frac{(E-E_{121})^2}{2\sigma_{121}^2}\right) \\
      &\qquad+ \frac{y_{121}}{1 + \exp\left(\frac{E-E_{121}}{\sigma_{121}/2}\right)},
  \end{split}
\end{align}
where the line energies $E_i$ and relative strengths $a_i$ in~\eqref{eq:f40} are given in Table~\ref{tab:f40lines}, $E_{121} = \SI{121.7817}{keV}$, the energy is related to ADC counts using the linear relation $E = (\mathrm{ADC} - b)/m$, and $b$, $m$, $C_{40}$, $A_{40}$, $\sigma_{40}$, $y_{40}$, $C_{121}$, $A_{121}$, $\sigma_{121}$, and $y_{121}$ are fit parameters.
In both functions, the onset of the Compton continuum due to scattering in the detector is approximated by the smooth ramp function with scales $y_{40}$ and $y_{121}$.
Each fit is performed twice, and during the second iteration, the fit ranges are set to $E_2 - 4\sigma_{40}$ to $E_4 + 4\sigma_{40}$ for $f_{40}$ and to $E_{121} \pm 4\sigma_{121}$ for $f_{121}$, where $\sigma_{40}$ and $\sigma_{121}$ are obtained from the first iteration.
An example fit is shown in Fig.~\ref{fig:spectrum28_0-detail}.
Typical values are $b \approx \SI{415}{ADC}$ and $m \approx \SI{2.9}{ADC/keV}$.
Figure~\ref{fig:resolutions} shows the distribution of energy resolutions at \SI{\sim 40}{keV} and \SI{121}{keV} of all active channels.
We find a median FWHM resolution of \SI{3.35}{keV} and \SI{3.60}{keV}, respectively.

\begin{table}
    \centering
    \caption{$^{152}\text{Sm}$ emission lines considered in the \SIrange{39}{46}{keV} line complex described by the function $f_{40}(E)$ defined in Eq.~\eqref{eq:f40} used to fit the calibration spectra of \xcalibur.}
    \begin{tabular}{ccSS}
    \toprule
      $i$ & Line(s)                         & {Energy $E_i/\si{keV}$}  & {Scale $a_i$} \\
    \midrule
      1   & $\text{K\textalpha}_1$           & 40.1186                  & 1             \\
      2   & $\text{K\textalpha}_2$           & 39.5229                  & 0.552         \\
      3   & $\text{K\textbeta}_{1,3,5''}$    & 45.4777                  & 0.312         \\
    \bottomrule
    \end{tabular}
    \label{tab:f40lines}
\end{table}

\begin{figure}
    \centering
    \includegraphics[width=\linewidth]{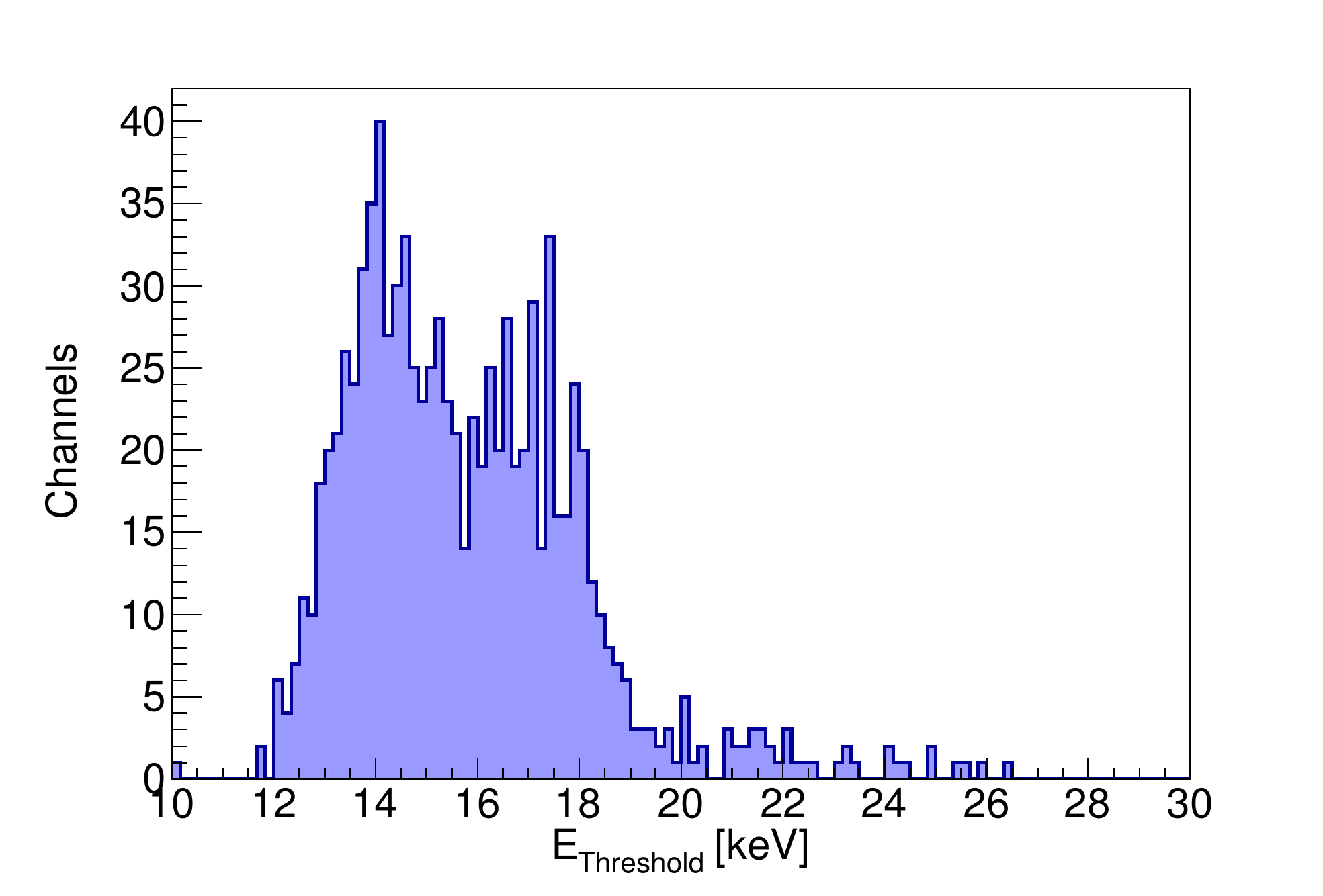}
    \caption{Distribution of energy thresholds defined as the \num{0.25}\textsuperscript{th} percentile of events in calibration data.}
    \label{fig:thresholds}
\end{figure}

Excessively noisy and non-operational channels were deactivated.
During the flight \num{949} of the 1088 channels (\SI{87}{\percent}) were working.
This number fell slightly short of the target of \SI{90}{\percent} working channels, and was accepted in order to achieve a stable detector operation.
Using the calibration data, we also determined the energy threshold of each channel.
Due to the hysteresis of the threshold discriminator in the ASIC, there is no sharp threshold.
Hence, we empirically defined the threshold as the \num{0.25}\textsuperscript{th} percentile of all events.
The distribution of energy thresholds is shown in Fig.~\ref{fig:thresholds}.

\subsection{Data acquisition and dead time}\label{sub:deadtime}
\begin{figure}
  \centering
  \includegraphics[width=\linewidth]{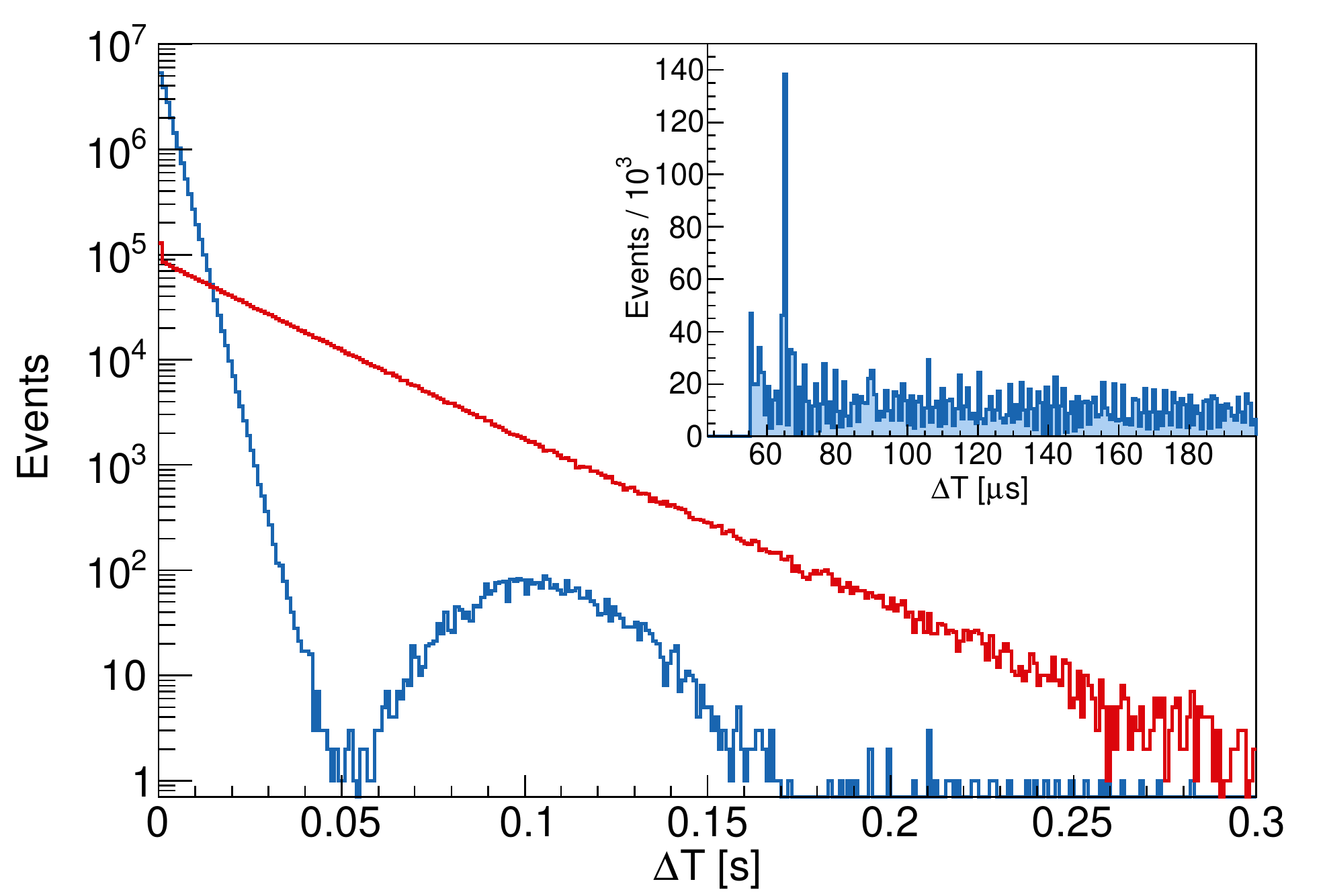}
  \caption{Time interval between consecutive events (blue: all events, red: events not vetoed by the anti-coincidence shield). The inset shows the distribution of small time intervals for all events.}
  \label{fig:deltaT}
\end{figure}
Figure~\ref{fig:deltaT} shows the time difference between consecutive events in \xcalibur throughout the entire flight.
From an exponential fit we find a rate of \SI{333}{Hz} for all triggers, and \SI{39}{Hz} for events not vetoed by the anti-coincidence shield.

\begin{figure}
  \centering
  \includegraphics[width=\linewidth]{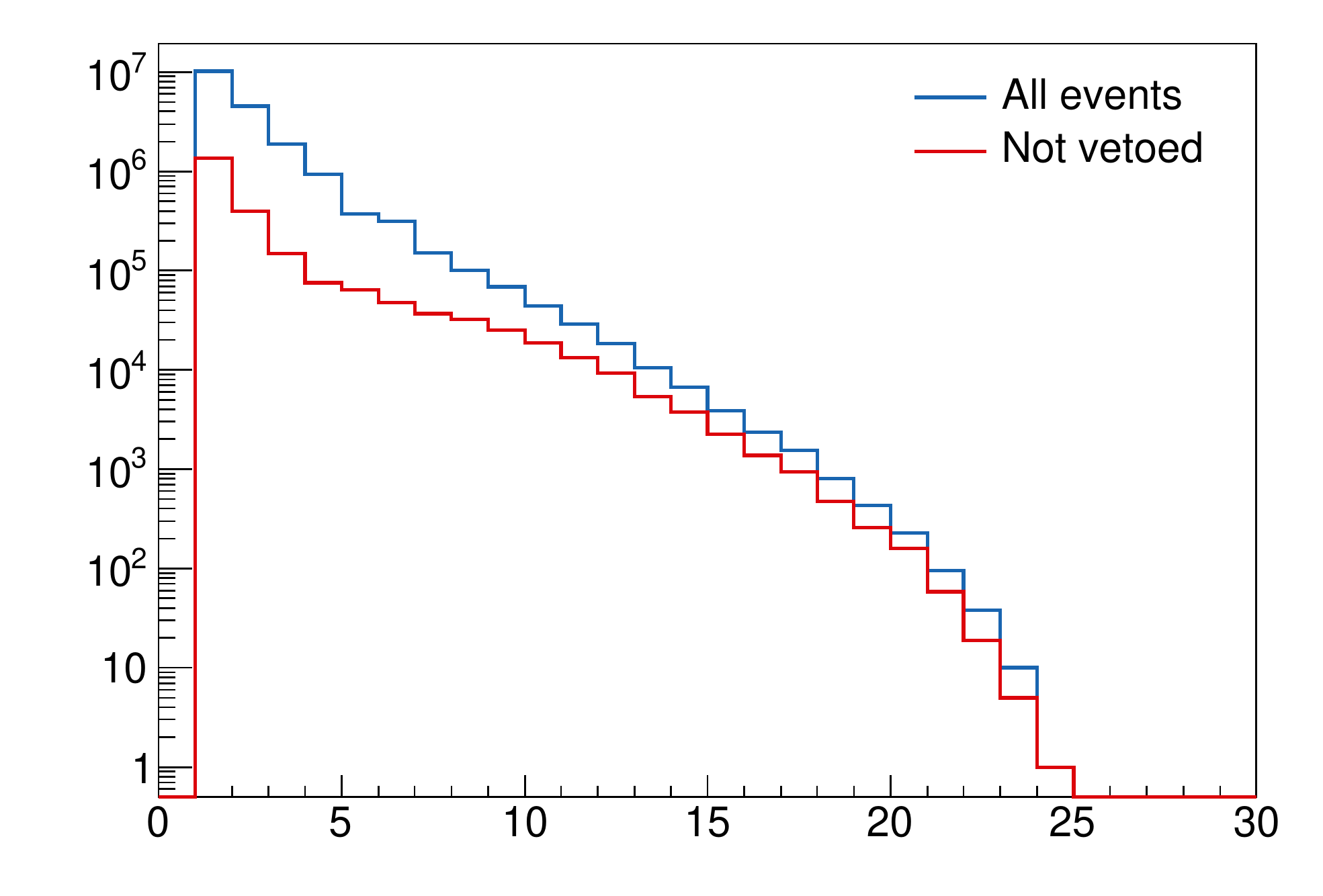}
  \caption{Channel multiplicity of events recorded during the flight. The theoretical maximum multiplicity is 24 since at most 3 channels from each ASIC will be read out, and data from at most 2 ASICs from each of the 4 FPGA boards can be buffered for readout. In the polarization and spectral analysis only single-channel events are considered.}
  \label{fig:multiplicity}
\end{figure}

As described in Section~\ref{sub:polarimeter}, the readout dead time for a single-pixel event is \SI{43.6}{\micro\second}.
From the multiplicity distribution shown in Fig.~\ref{fig:multiplicity}, we estimate an average dead time of \SI{52}{\micro\second}.
During this dead time, no events will be recorded.
Given the trigger rate at float, this corresponds to \SI{1.8}{\percent} of events.
However, due to the inability of the firmware to distinguish between multiple ASICs read as part of the same trigger or of consecutive triggers, we decided to discard all events with a time difference less than \SI{55}{\micro\second}, resulting in an artificial dead time of \SI{1.9}{\percent}.
This artificial dead time can be seen in the inset of Fig.~\ref{fig:deltaT} and does not add to the detector readout dead time.
The spike at \SI{65}{\micro\second} is currently not fully understood, but likely due to the way subsequent events are time-tagged after being transmitted from the front-end readout to the CPU.

The broad bump at $\SI{0.05}{s} < \Delta T < \SI{0.15}{s}$ in the distribution of all events in Fig.~\ref{fig:deltaT} is due to individual lost data packets.
These packets have been lost during transmission through the rotating connector.
We are implementing improvements to this link to prevent such data loss during the follow-up mission \xlcalibur.

\subsection{Performance of the anti-coincidence shield}\label{sub:shield}
The anti-coincidence shield performance was tested after the flight with a $^{137}\text{Cs}$ radioactive source, which emits \SI{662}{keV} gamma-rays.
Since the flight readout board does not have a pulse-height digitization capability, we used an external pre-amplifier (CANBERRA 2005), shaping amplifier (ORTEC 485) and multi-channel analyzer (Amptek 8500D) for the spectral and gain studies.
During the flight, we set the PMT high-voltage to \SI{750}{V} for all PMTs and the veto threshold to \num{30} and \num{300} channels for the PMTs connected to the top and bottom part of the shield, respectively.
For the bottom part, we were able to set the threshold as low as \num{30} channel, but increased to \num{300} due to the high rate of cosmic-rays during the flight, which would have resulted in an unacceptably large dead time.

\begin{figure*}
  \centering
  ~\hfill%
  \includegraphics[width=.4\textwidth]{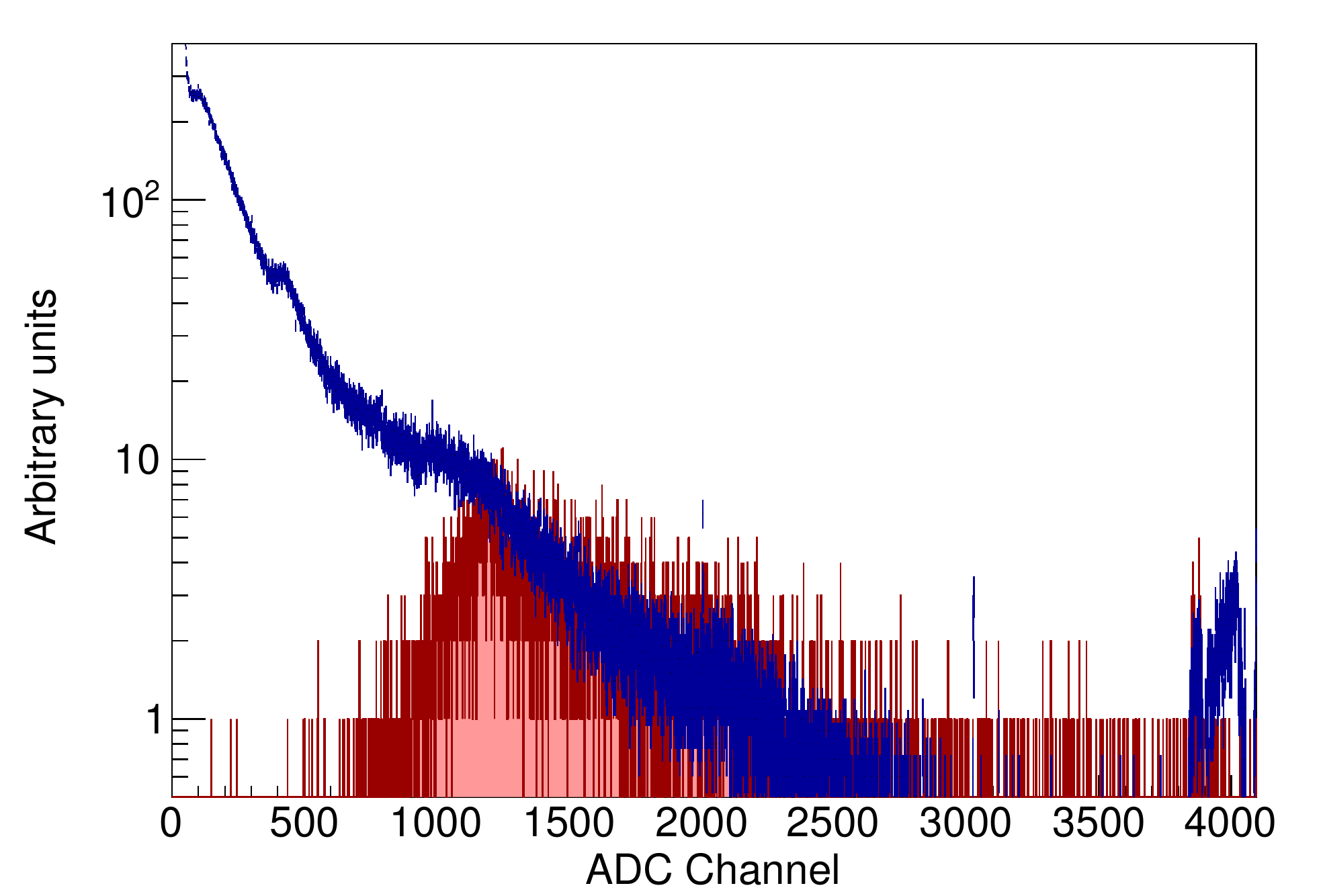}%
  \hfill%
  \includegraphics[width=.4\textwidth]{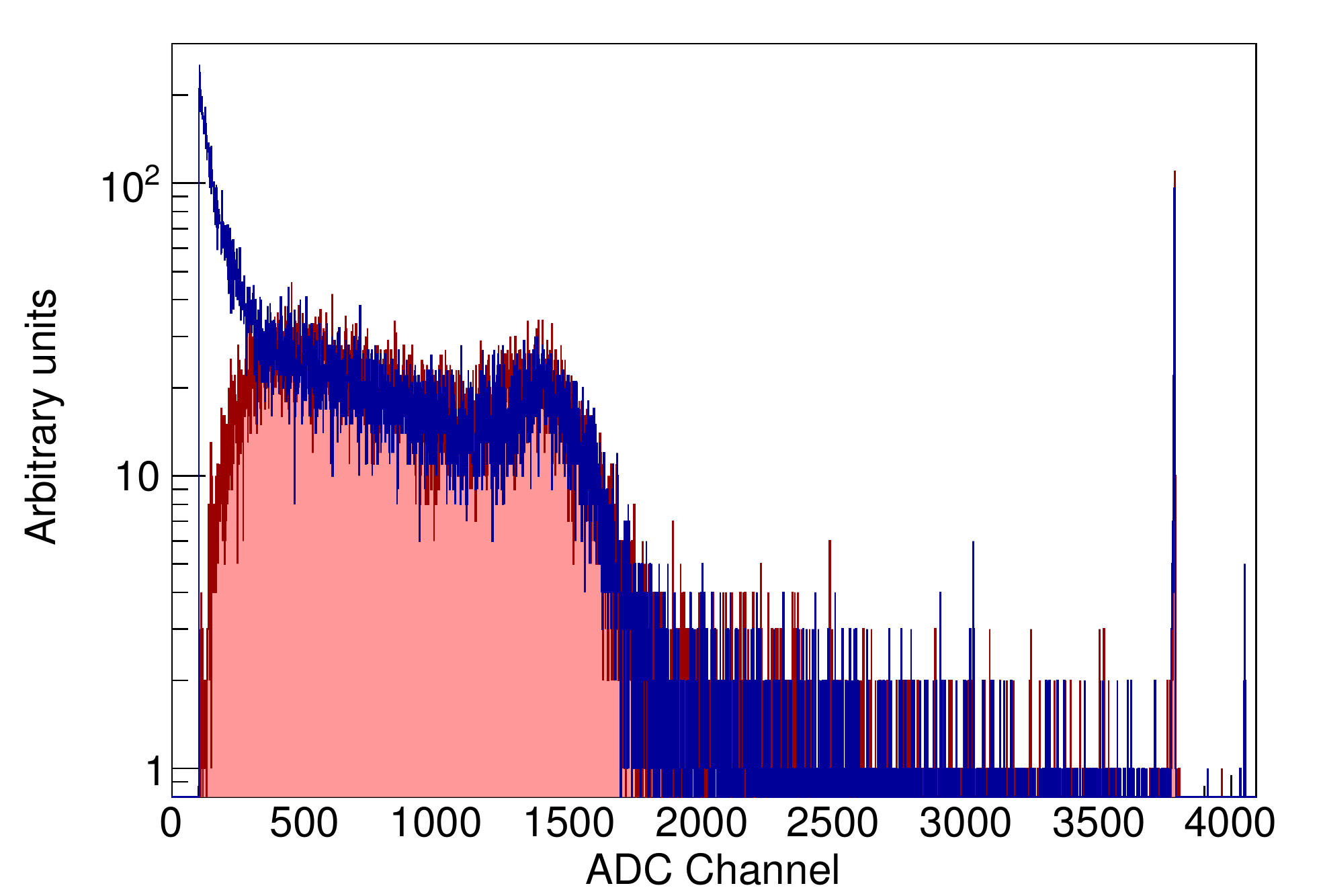}%
  \hfill~%
  \caption{\emph{Left:} $^{137}\text{Cs}$ spectra obtained by the bottom part of the anti-coincidence shield. The blue histogram is measured with the self-trigger of the external multi-channel analyzer (i.e., without the trigger from the flight readout board). It shows \SI{662}{keV} photoabsorption peak at \num{\sim 500} channel. The red histogram is obtained using the trigger from the flight board as the external trigger. The exposure of the two measurements is not the same, and spectra have been scaled to match count rates above the veto threshold. \emph{Right:} Same for the top part of the anti-coincidence shield. The \SI{662}{keV} peak is at \num{\sim 1400} channel.}
  \label{fig:shield-spectra}
\end{figure*}

With the flight operation high-voltage of \SI{750}{V}, we obtained $^{137}\text{Cs}$ spectra for each PMT and confirmed that the gain is consistent within a factor of \num{\sim1.5} within each shield part.
In order to measure the trigger threshold, the signal from one PMT is input to the external spectral measurement, and a trigger signal is generated from the other three PMTs using the flight electronics.
The spectra of both top and bottom parts are obtained with (red) and without (blue) the trigger from the flight board and shown in Fig.~\ref{fig:shield-spectra}.
For the bottom, the flight board spectrum shows a cut-off around \num{1200} channel which corresponds to \SI{\sim 1.4}{MeV}.
Since the veto comparator checks the signal after summing all 4 PMTs (we use only 3 PMT signals for the trigger in this measurement), we estimate the flight veto threshold for the 2018/19 flight at $\SI{1.4}{MeV} / (4/3) \sim \SI{1.1}{MeV}$.
In case of the top part, the cut-off appears at \num{\sim 400} channel corresponding to \SI{\sim 200}{keV}, an estimated flight veto threshold of \SI{\sim 150}{keV}.
The high threshold of the bottom shield was necessary in order to keep the dead time below~\SI{5}{\percent} (see Section~\ref{sec:background} and Fig.~\ref{fig:scaler}).

These measurements show that it will be possible to significantly reduce the threshold of the bottom part of the CsI shield.
The \SI{6.4}{\micro\second} duration of the veto signal was necessitated by the time constant of the amplifier.
For future flights of \xlcalibur, we will modify the analog circuit of the readout board to reduce the time constant of the amplifier.
Furthermore, we will tune the duration and timing of the digital anti-coincidence flag in order to reduce the dead time and maximize veto efficiency.
Finally, the \xlcalibur shield will be made of BGO, which enables a more compact geometry due to its higher stopping power and inherently provides faster scintillation signals~\cite{abarr_etal_2021}.

\section{Background}\label{sec:background}
\begin{figure}
    \centering
    \includegraphics[width=\linewidth]{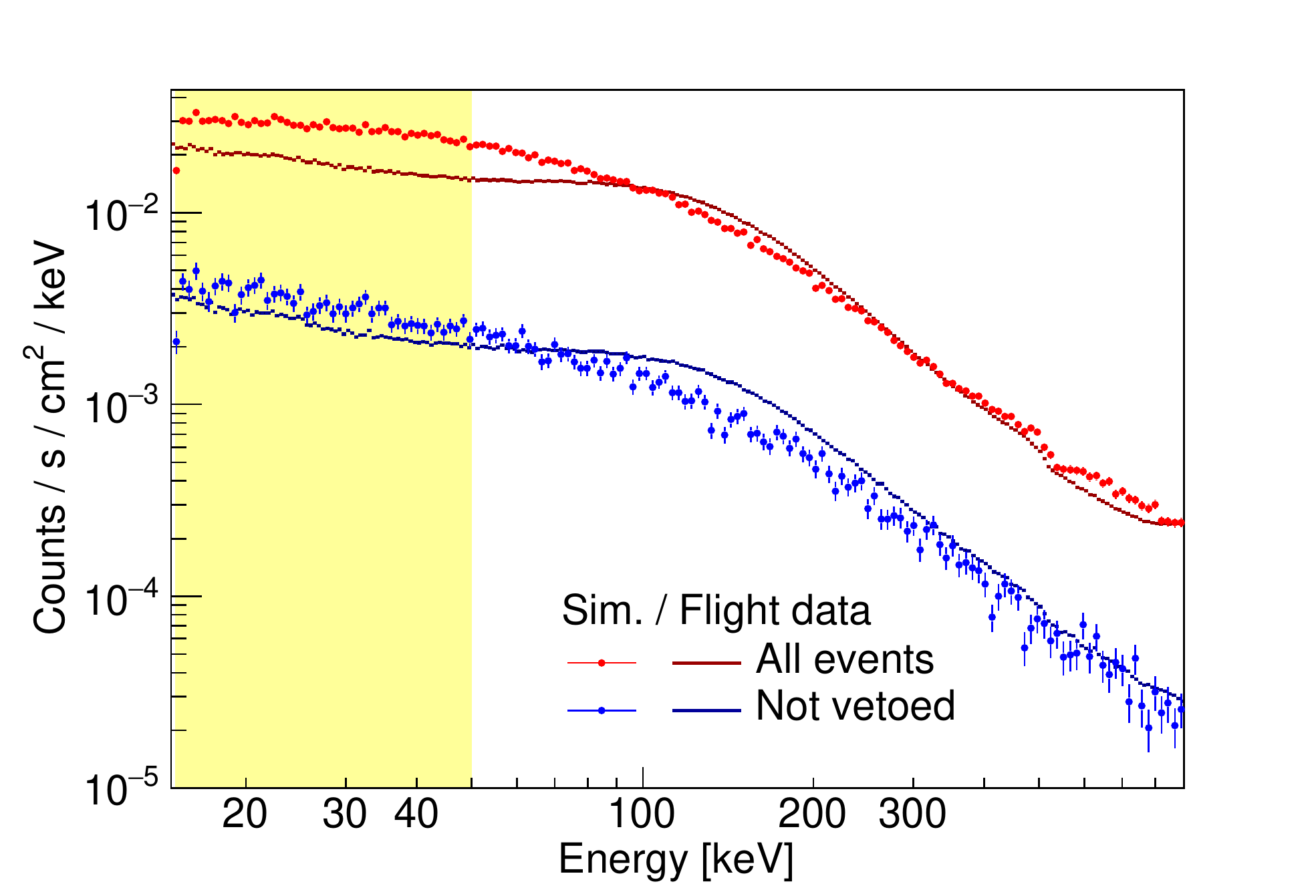}
    \caption{Background spectrum measured during flight, excluding the ASIC of the rear-end detector that only worked intermittently, and the result of the Geant4 simulation. The shaded region indicates the core energy range for polarization measurements.}
    \label{fig:background-spectrum}
\end{figure}

As described in Section~\ref{sub:wasp}, source observations were interspersed with frequent off-source pointings to measure the background rates and spectrum in the detector.
In total, \num{1.6e7} events were recorded during a detector live time of \SI{94}{ks} for background observations after reaching floating altitude.
The overall single-pixel background spectrum, compared to simulations is shown in Fig.~\ref{fig:background-spectrum}, both for all events
and selecting only those events not vetoed by the anti-coincidence shield.
In the energy range of interest for polarization measurements (\SIrange[range-phrase={ to }]{15}{50}{keV}), the background rate for single-pixel events after subtracting events vetoed by the anti-coincidence shield was \SI{2.1}{Hz}.

\begin{figure}
  \centering
  \includegraphics[width=\linewidth]{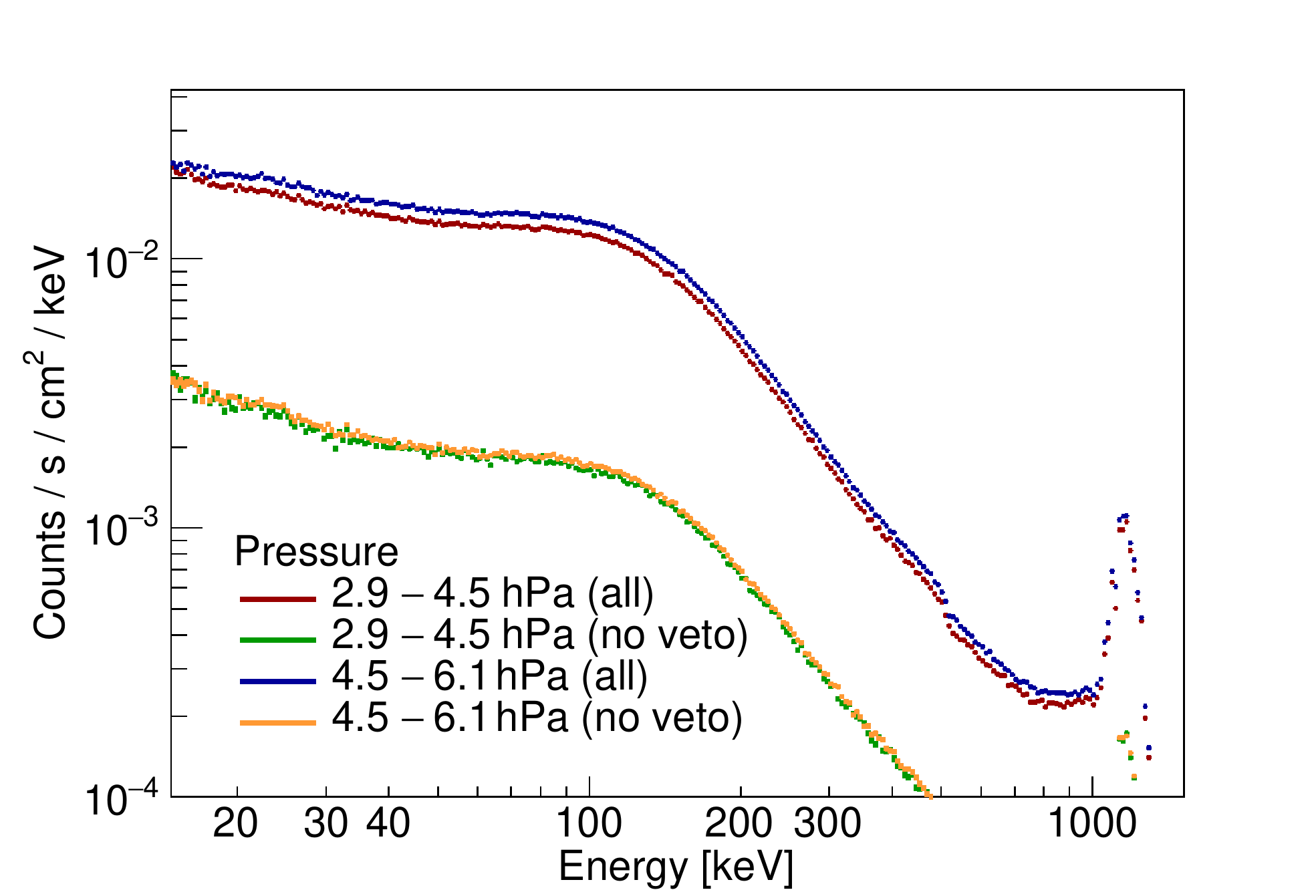}
  \caption{Background spectrum for two different balloon pressure altitude ranges.}
  \label{fig:background-by-pressure}
\end{figure}

\begin{figure}
  \centering
  \includegraphics[width=\linewidth]{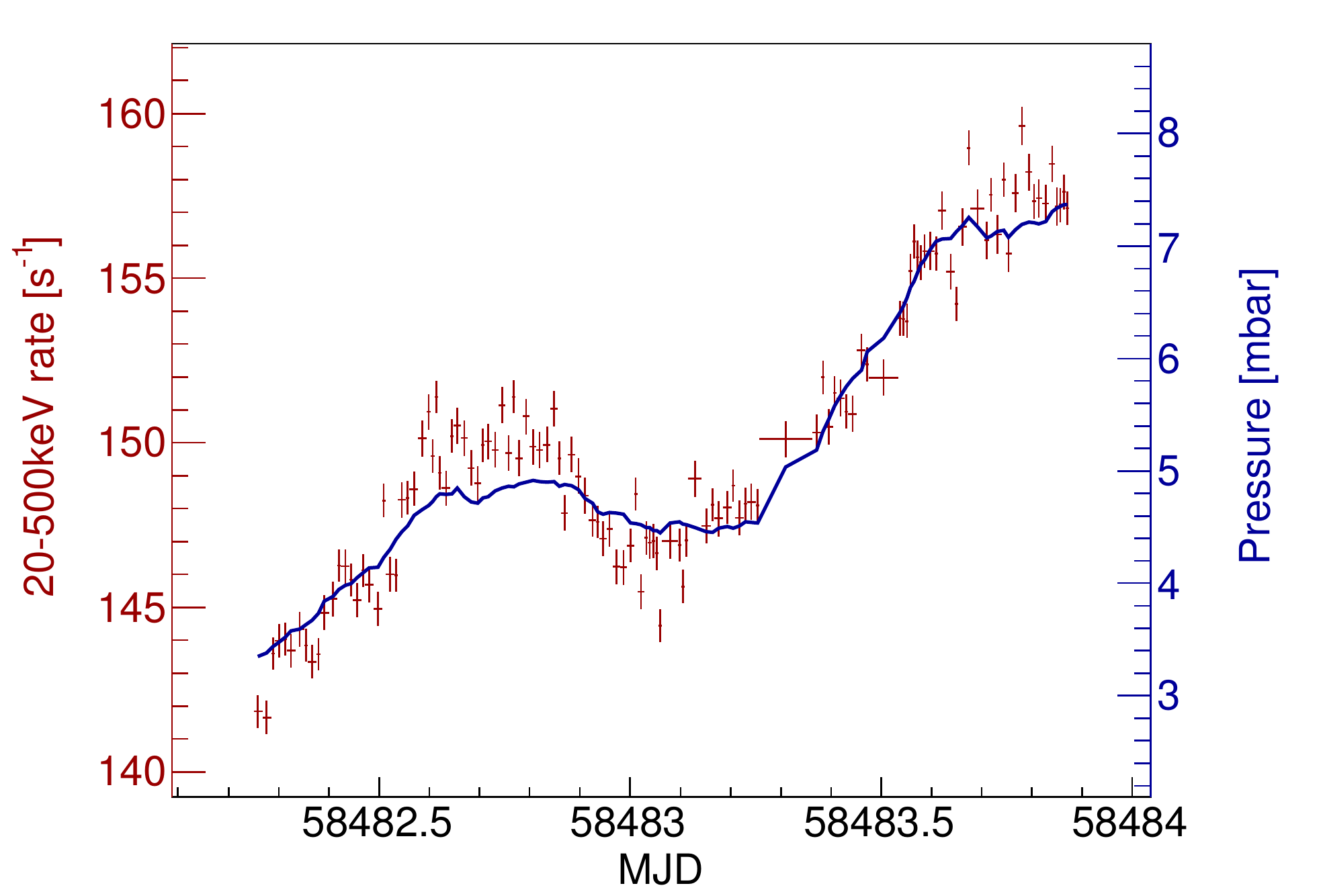}
  \caption{Correlation between event rate and pressure altitude during background observations. The figure shows the rate of all single-pixel events.}
  \label{fig:rate_v_pressure}
\end{figure}

Figure~\ref{fig:background-by-pressure} shows that at higher altitudes (lower pressure) the background rate is lower, and that the effect is more pronounced before application of the anti-coincidence veto.
The spectral shape barely changes.
The strong correlation of background rate with pressure is also illustrated in Fig.~\ref{fig:rate_v_pressure}.

\begin{figure}
  \centering
  \includegraphics[width=\linewidth]{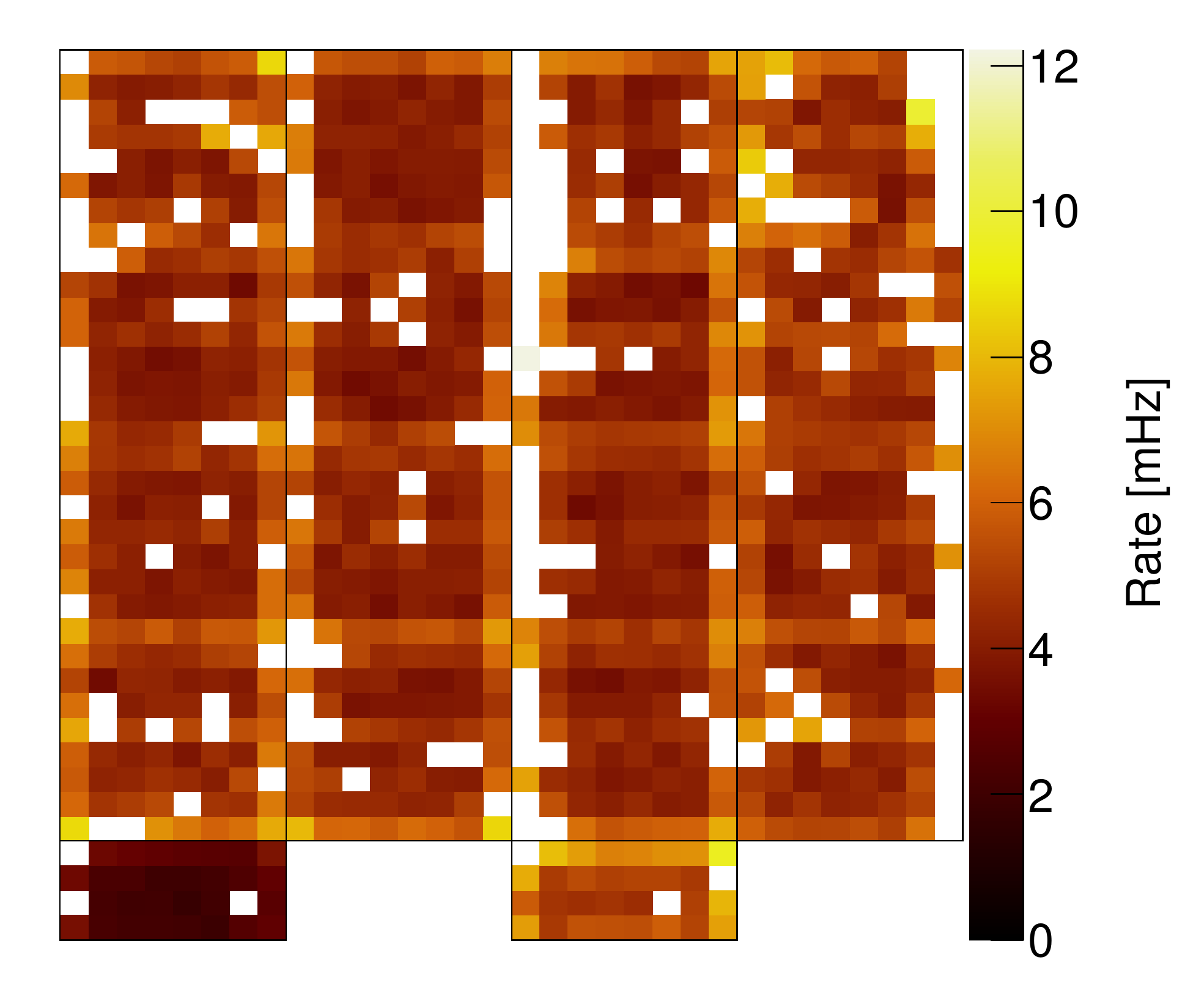}
  \caption{\SIrange{25}{60}{keV} background event rate in all pixels of events not vetoed by the anti-coincidence shield. The four sides of the polarimeter are laid out next to each other and the two ASICs reading out the \nth{17} CZT are shown on the bottom below their corresponding readout FPGA boards. Background rates are generally higher at the edges of CZT detectors where particles entering the detector from the side can trigger the readout.}
  \label{fig:pixelmap}
\end{figure}

Figure~\ref{fig:pixelmap} shows the fairly uniform distribution of background events in the detector.
It also shows that the majority of dead or disabled pixels are on the edges of detectors.
This is not entirely unexpected as detector edges are particularly vulnerable to leakage currents, which result in a high noise level in the front end.
Furthermore, pixels along the edges of CZTs, particularly along the board edges have a higher background rate due to their larger surface area.
The azimuth distribution with respect to celestial north shows a~\SI{\sim 0.2}{\percent} modulation.
The origin of this modulation is unknown, but we speculate that it may be a residual effect of an up/down asymmetry of the background, or due to a slow evolution of the background that does not average out with the rotations.
It should be noted that this effect is small compared to other potential sources of systematic error.

\begin{figure}
  \centering
  \includegraphics[width=\linewidth]{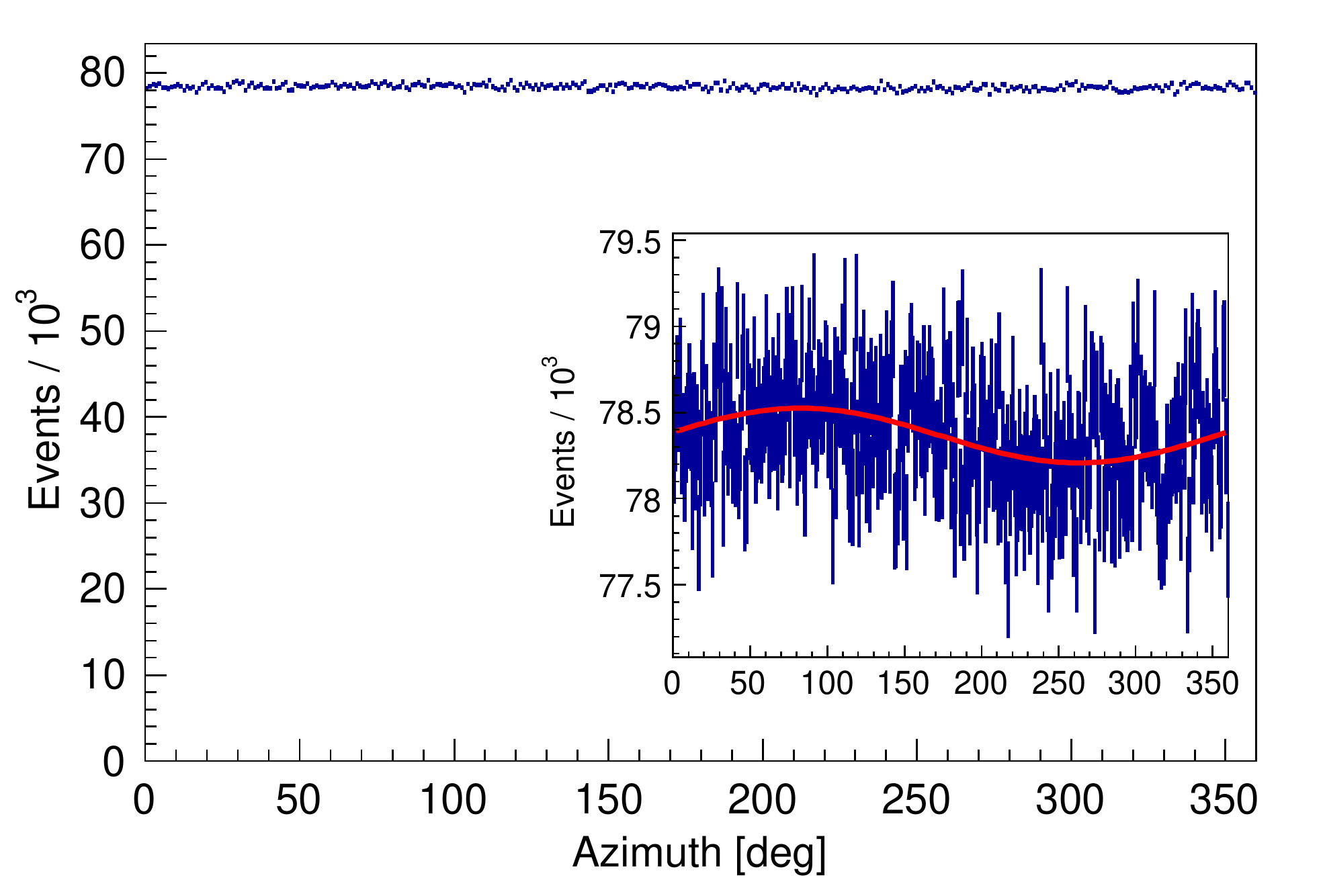}
  \caption{Azimuth distribution of background events with respect to celestial north. The inset shows a zoom into the same distribution.
  The sinusoidal fit reveals a residual modulation of~\SI{\sim 0.2}{\percent}.}
  \label{fig:background_azimuth}
\end{figure}

One half of the rear-end CZT, shown on the bottom of the first column of detectors, has an anomalously low rate.
The readout ASIC for this part of the detector worked intermittently about \SI{50}{\percent} of the time.
While working, the data from this ASIC were good by all our metrics, in particular the measured background spectrum.
In the analysis, the lifetime of this ASIC is taken into account separately based on the returned data.
No definite cause of the issue, which appeared after launch, has been determined and no other ASIC showed similar behavior.

\begin{figure*}
    \centering
    \includegraphics[width=0.47\linewidth]{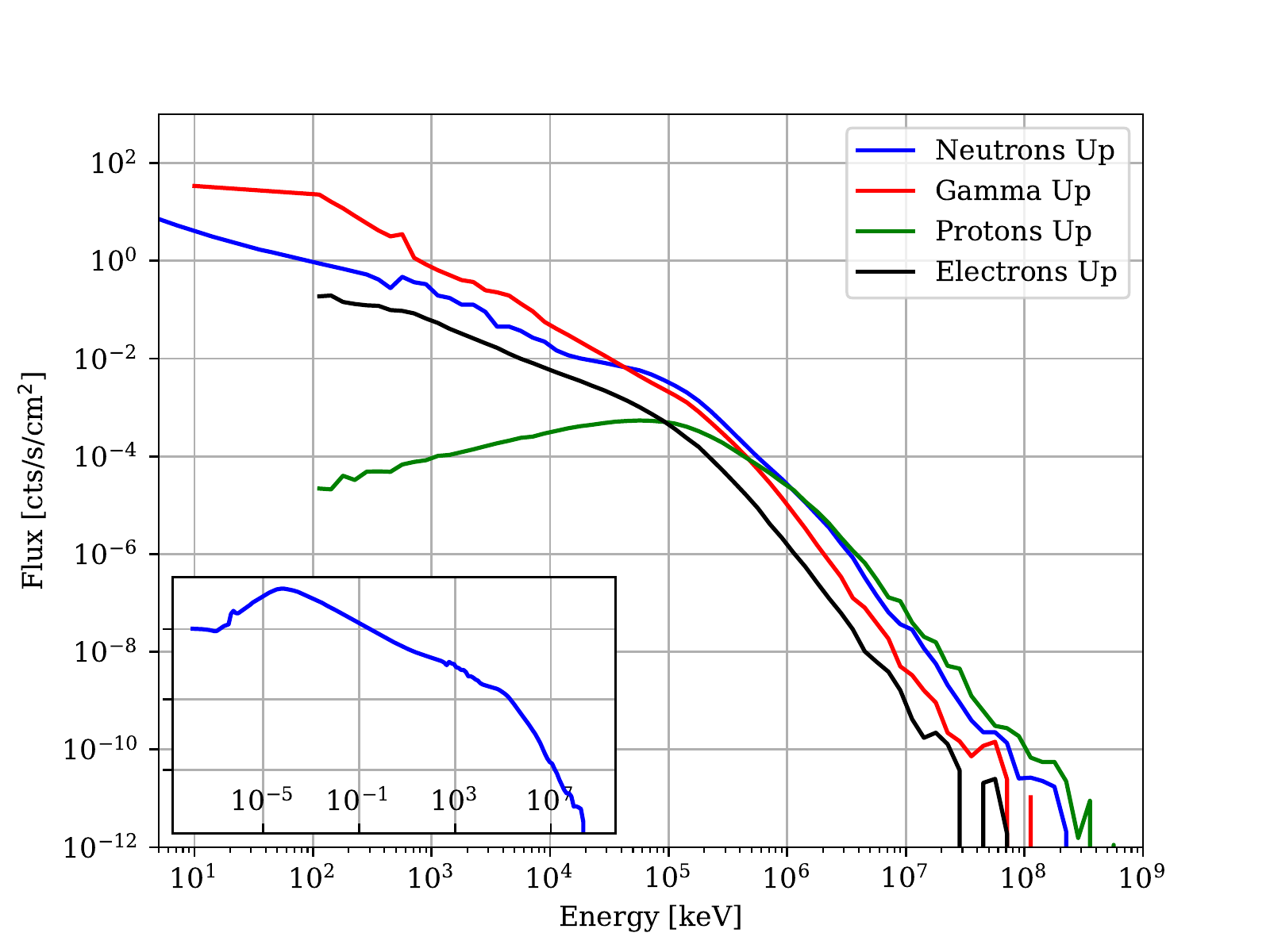}
    \includegraphics[width=0.47\linewidth]{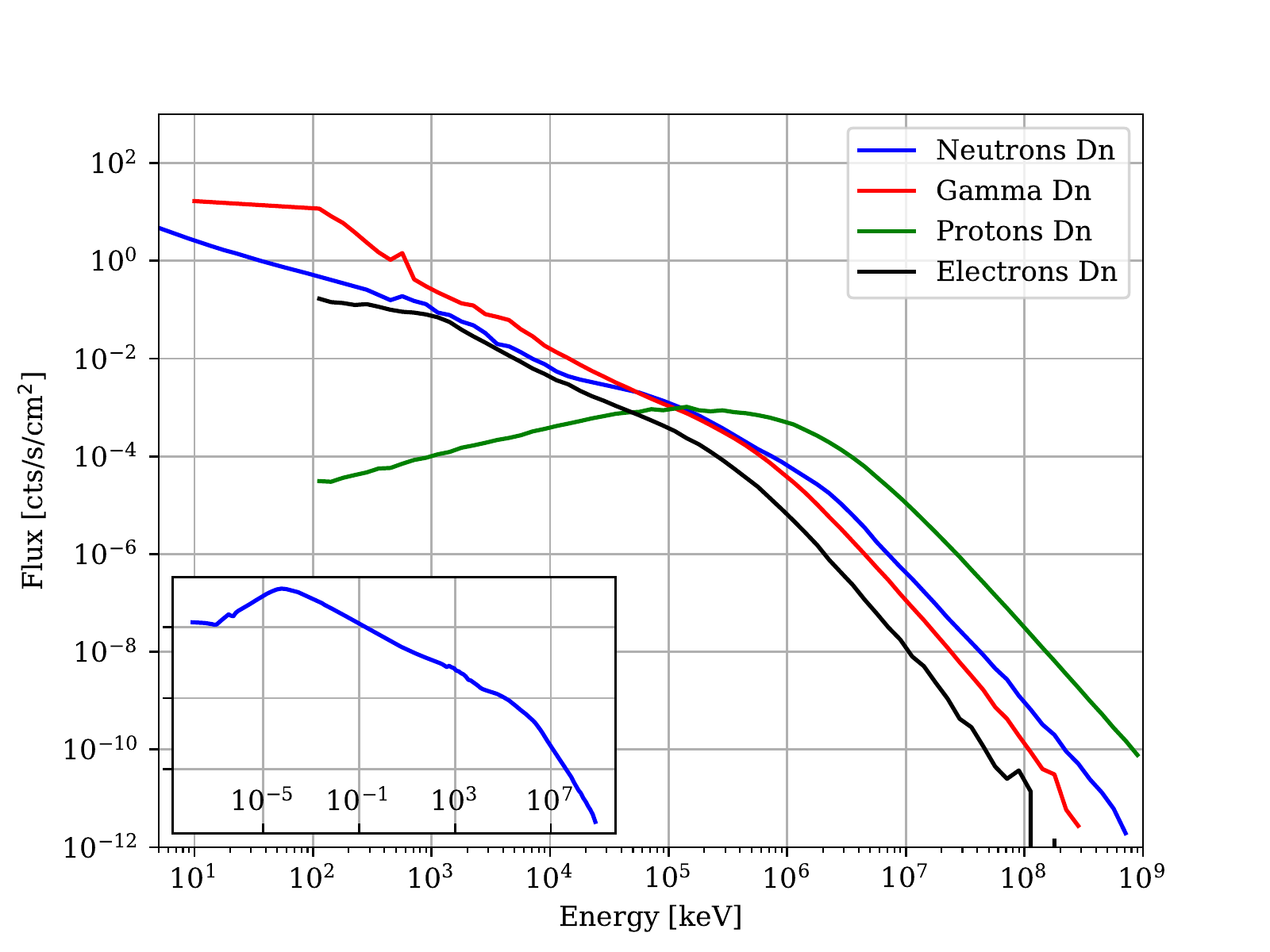}
    \caption{Upward and downward moving particle spectra as obtained from MAIRE. Inset shows the neutron spectrum down to $1\mu\text{eV}$.}%
    \label{fig:maire}
\end{figure*}

Figure~\ref{fig:background-spectrum} also shows the result of a simulation in comparison with the measured background spectrum.
The polarimeter and anti-coincidence shield geometry were modeled in Geant4~\cite{agostinelli_etal_nima_2003,allison_etal_ieee_2006,allison_etal_nima_2016}, and incident particles were sampled across the upper and lower (albedo) hemispheres with spectra from MAIRE~\cite{2006ITNS...53.1851L} shown in Fig.~\ref{fig:maire}.
The simulations are described in more detail in Ref.~\cite{abarr_etal_2021}.

\begin{figure}
    \centering
    \includegraphics[width=\linewidth]{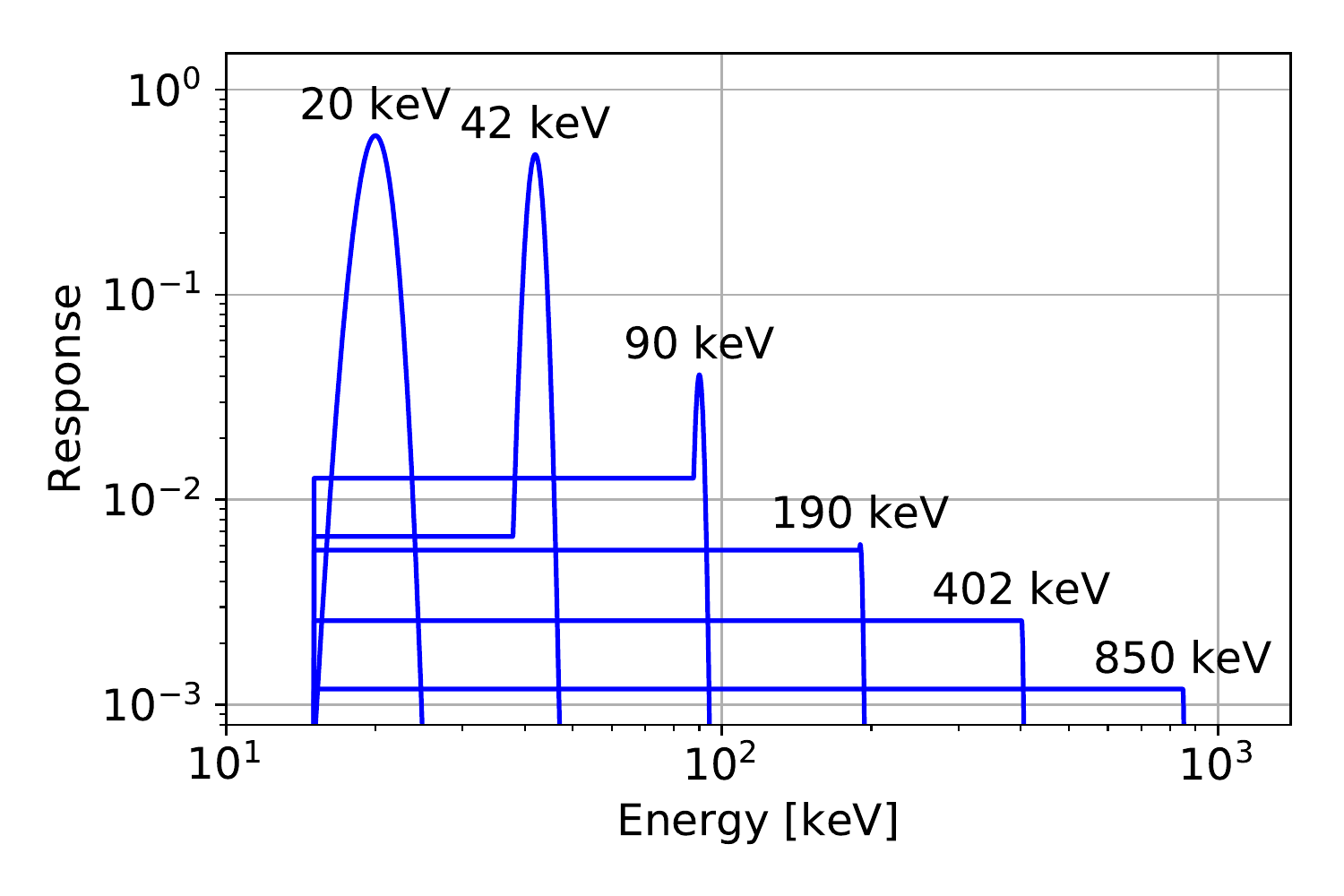}
    \caption{CZT energy redistribution function composed of a Gaussian and a tail as used in the background simulations.}%
    \label{fig:cztresp}
\end{figure}

The energy response of the CZT detectors is applied to the output from Geant4.
It is approximated using an asymmetric Gaussian distribution with a flat tail to low energies, as shown in Figure~\ref{fig:cztresp}.
This response reflects the detector calibrations using $^{152}\text{Eu}$ (see Section~\ref{sub:calibration} and Ref.~\citep{beilicke_etal_jai_2014}). 
A phenomenological low-energy tail was added to the detector response, bringing the observed and simulated energy spectra into agreement.
Simulations of the detector response based on Geant4 simulations
of the interactions of primaries and secondaries in the detector substrate
and simulations of charges drifting through the detector 
do not fully explain the low-energy tail. 
The effect is assumed to be the same for all particle species. The anti-coincidence shield was modeled by tagging events with an energy deposit \SI{>1}{MeV}, consistent with the measurements in Section~\ref{sub:shield}.

\begin{figure}
    \centering
    \includegraphics[width=\linewidth]{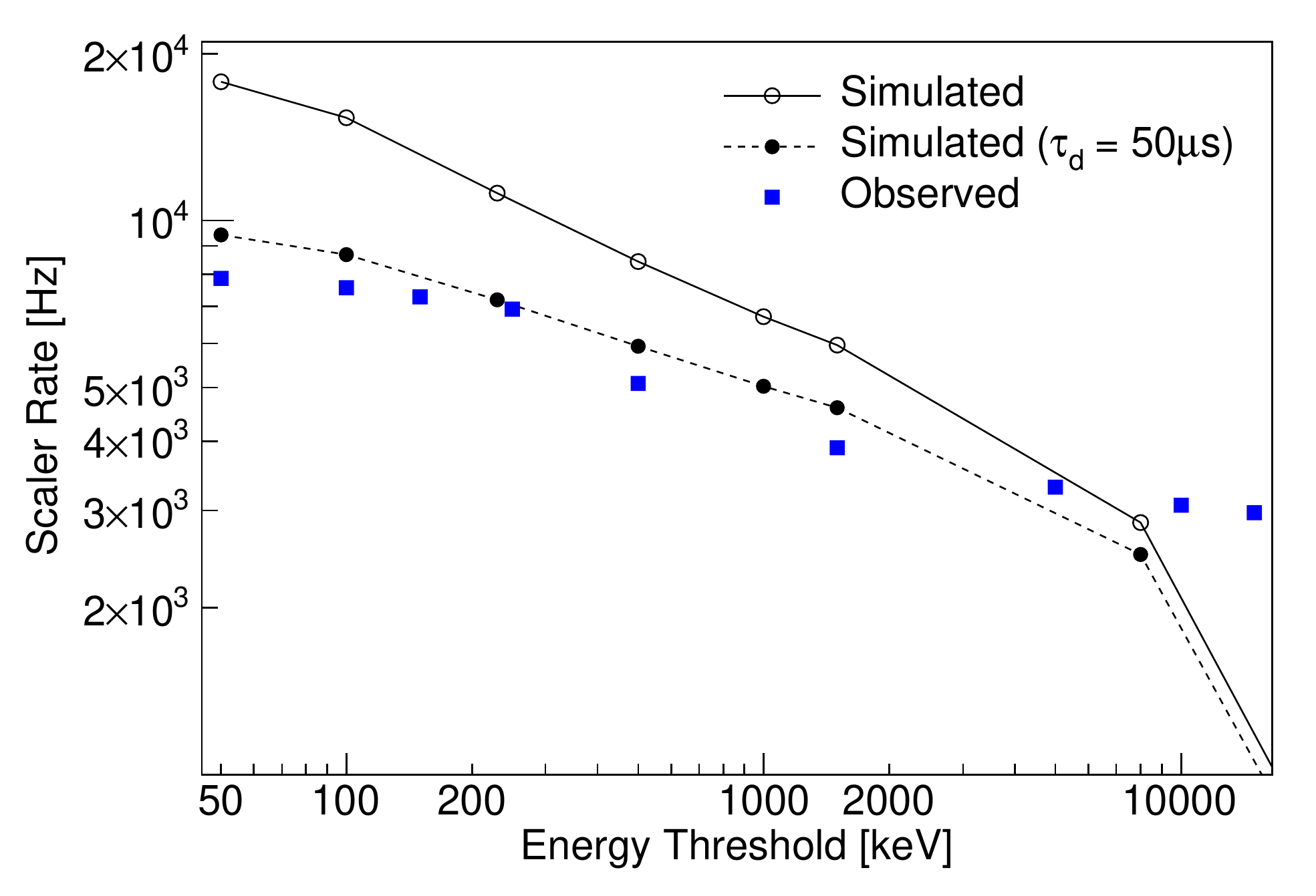}
    \caption{Scaler rates of the anti-coincidence shield as a function of trigger threshold measured during flight, compared to the results of a Geant4 simulation. A good agreement is achieved when assuming a deadtime of \SI{50}{\micro\second} per trigger in the shield.}
    \label{fig:scaler}
\end{figure}

\begin{figure}
  \centering
  \includegraphics[width=\linewidth]{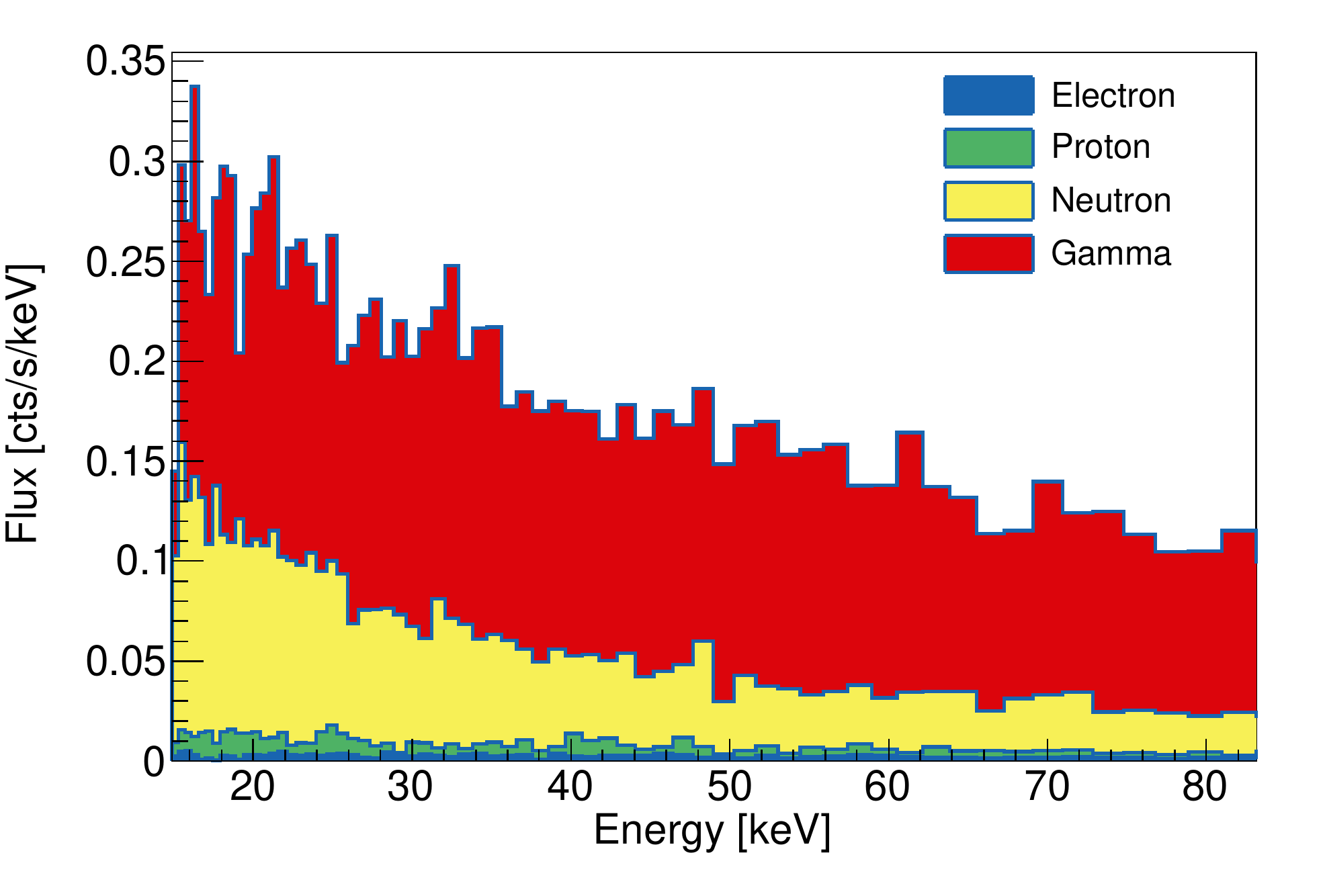}
  \caption{Contributions to the measured background by the different particle species shown in Fig.~\ref{fig:maire}, in the energy range of interest for polarization measurements after applying the anti-coincidence shield veto.}
  \label{fig:background_contributions}
\end{figure}

As Fig.~\ref{fig:background-spectrum} shows, we achieve a qualitative match between the simulation results and the measured background spectrum.
A comparison between observed and simulated trigger rates of the anti-coincidence shield accounting for a \SI{50}{\micro\second} dead-time for \SI{9}{MeV} minimum ionizing particles is shown in right panel of Figure~\ref{fig:scaler}.
After reaching floating altitude, we evaluated the response of the anti-coincidence shield by measuring the trigger rate as a function of trigger threshold.
The threshold DAC setting used during flight was converted to an energy threshold based on the post-flight measurements discussed in Section~\ref{sub:shield}.
The simulations reproduce the observations reasonably well, which gives confidence in using the simulation to study improvements in the polarimeter design.
The relative contributions of particle fluxes to the overall background count rate is shown in Figure~\ref{fig:background_contributions}.
As seen, gamma rays dominate the background in the energy range of interest (\SIrange[range-phrase={ to }]{15}{50}{keV}).
However, at low energies, neutrons also contribute significantly.
Further reduction of the neutron background could be achieved with a polyethylene neutron moderator, which has been studied for the follow-up mission \xlcalibur.
However, this would result in a significant increase in instrument mass resulting in increased mechanical complexity and potentially a lower flight altitude~\cite{abarr_etal_2021}.

\section{Detector Response}\label{sec:response}
We use Geant4 simulations of the \xcalibur detector~\cite{kislat_etal_aph_2015a} in order to determine the response of the detector to incident photons from the source.
For signal simulations, we use a simplified geometry consisting only of the scattering element and the CZT detectors.
The focal spot is centered on the axis of the scattering element and photons propagating parallel to the axis, distributed according to the mirror point spread function.
We simulated both unpolarized photons and photons that are \SI{100}{\percent} polarized.
Rotation of the detector is simulated by randomly changing the azimuthal orientation of the detector for each simulated event.\footnote{This should be a reasonable approximation of the uniform rotation of \xcalibur, which rotates \numrange{30}{45} times during each on-source observation, depending on the source.}
The response of the CZT detectors is modeled based on a parametrization of a simple charge tracking code, and then folded with the measured energy resolution for each pixel.
The energy threshold of each pixel is applied and hits in disabled pixels are removed.
The simulated data were processed in the same way as the experimental data~\cite{abarr_etal_2020}.

\begin{figure*}
  \centering
  ~\hfill%
  \includegraphics[width=.47\linewidth]{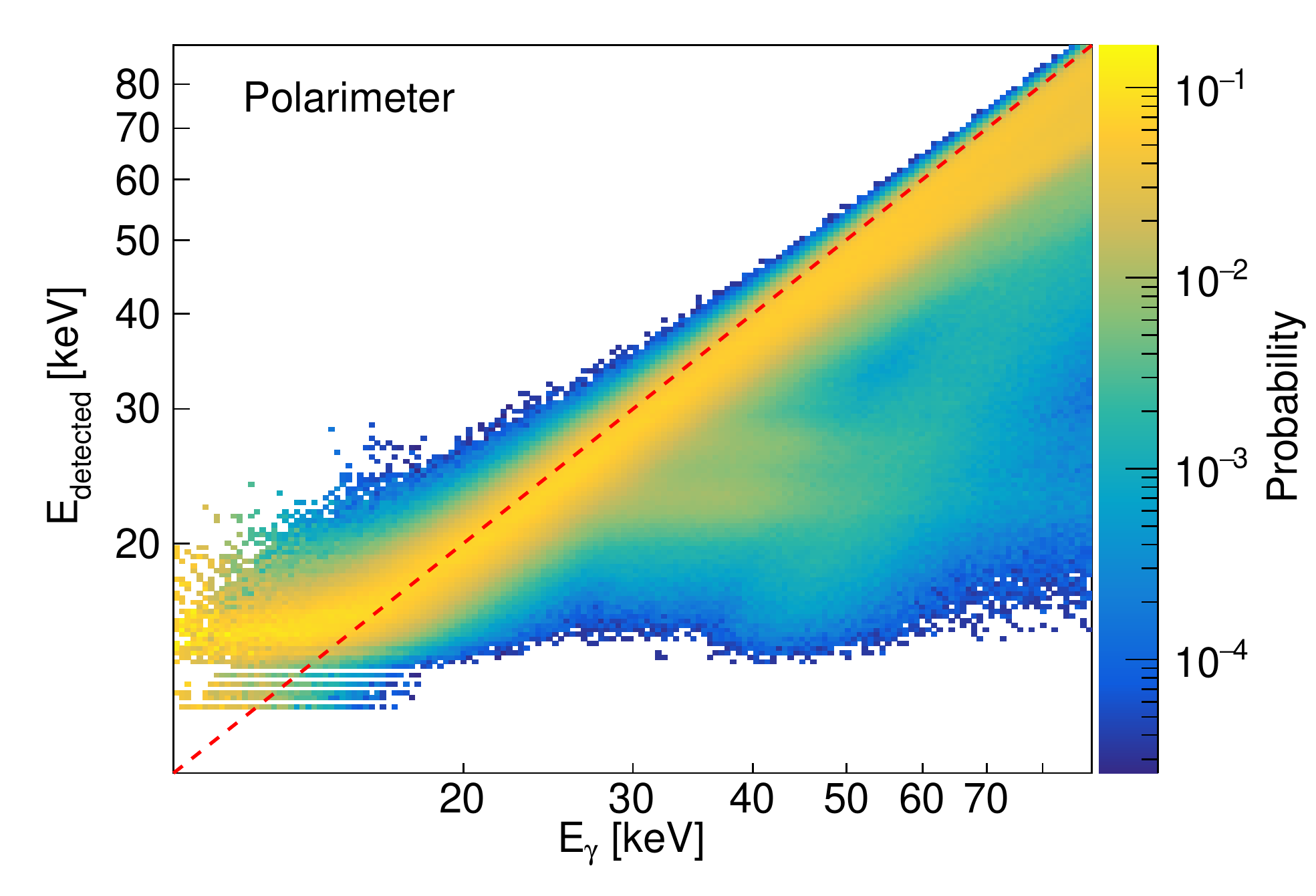}%
  \hfill%
  \includegraphics[width=.47\linewidth]{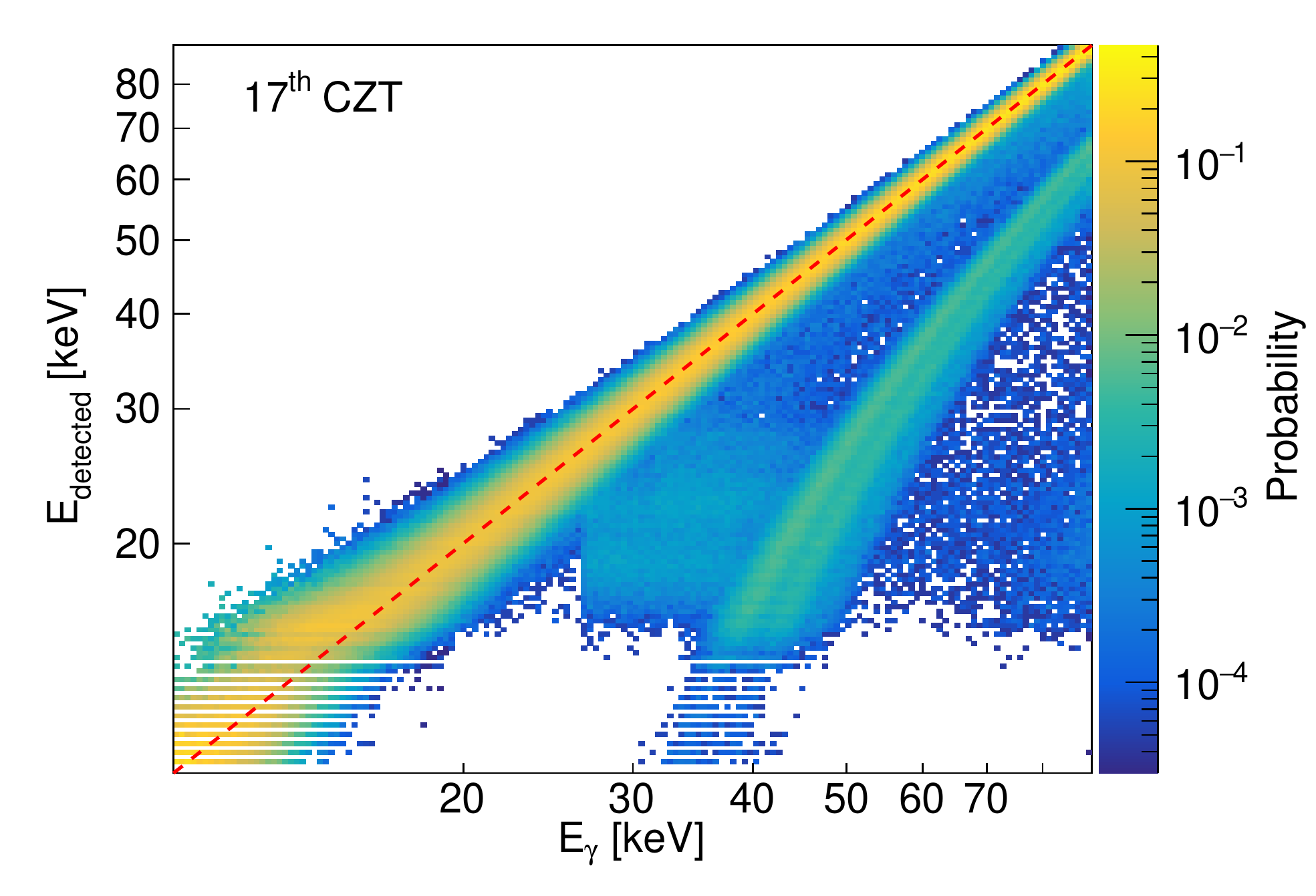}%
  \hfill~%
  \caption{\textit{Left:} Energy redistribution matrix for the 16 CZT detectors of the polarimeter. Each column is normalized so that color represents the probability to measure $E_\text{detected}$ given incident photon energy $E_\gamma$. The red dashed line indicates the diagonal where $E_\text{detected} = E_\gamma$. The bias towards $E_\text{detected} > E_\gamma$ at low energies is due to the trigger threshold of the detectors and appears significant due the column-wise normalization of the figure. \textit{Right:} Same for the \nth{17} CZT detector. The energy resolution appears better because most photons in this detector did not scatter. The band below the diagonal is due to escape peaks (see Fig.~\ref{fig:spectrum28_0-detail}). In the left figure it is significantly broadened because the energy of the scattered photon impinging on the CZT detectors differs from the incident photon energy $E_\gamma$ depending on the scattering angle.}
  \label{fig:ResponseMatrix}
\end{figure*}

\begin{figure*}
  \centering
  ~\hfill%
  \includegraphics[width=.47\linewidth]{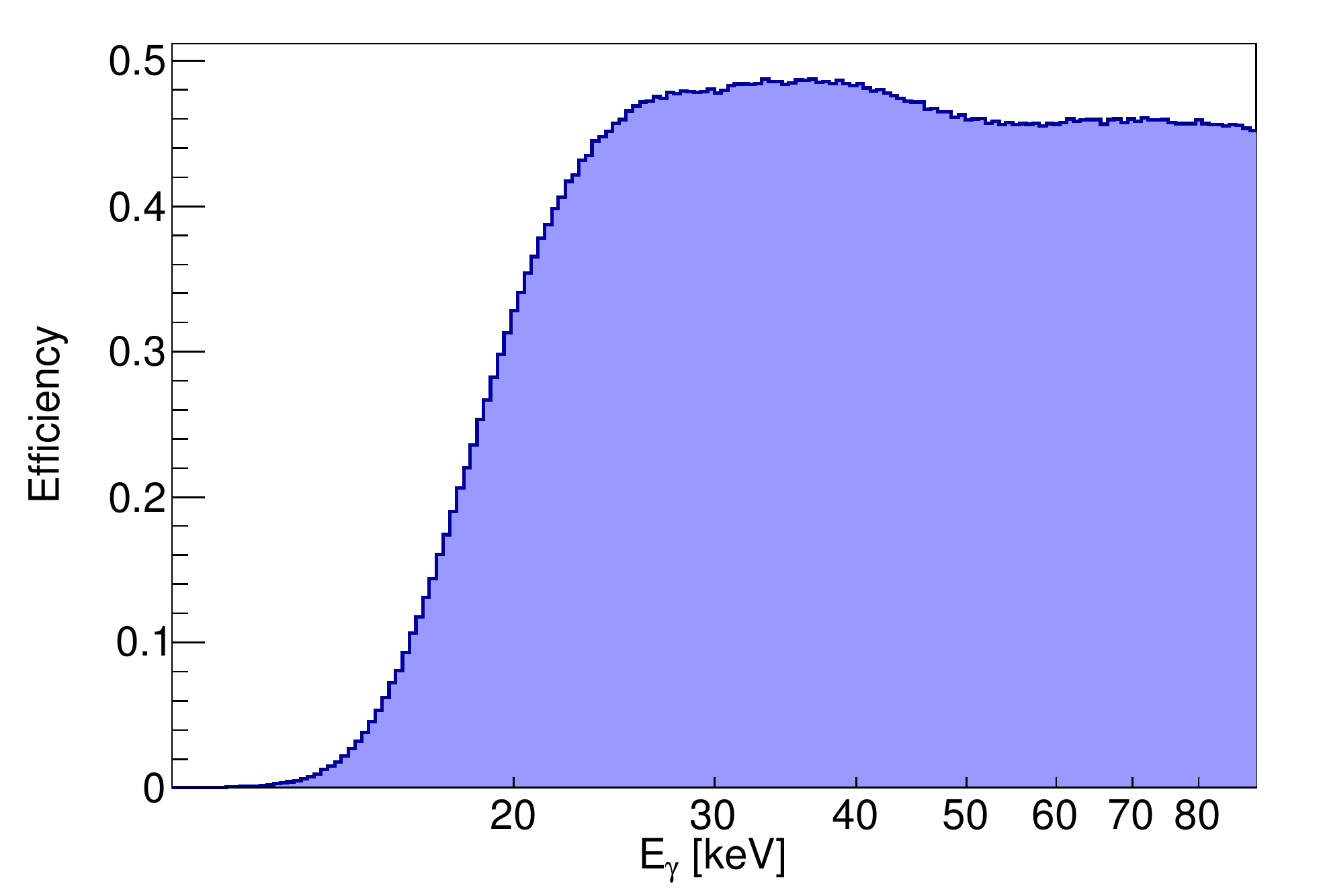}%
  \hfill%
  \includegraphics[width=.47\linewidth]{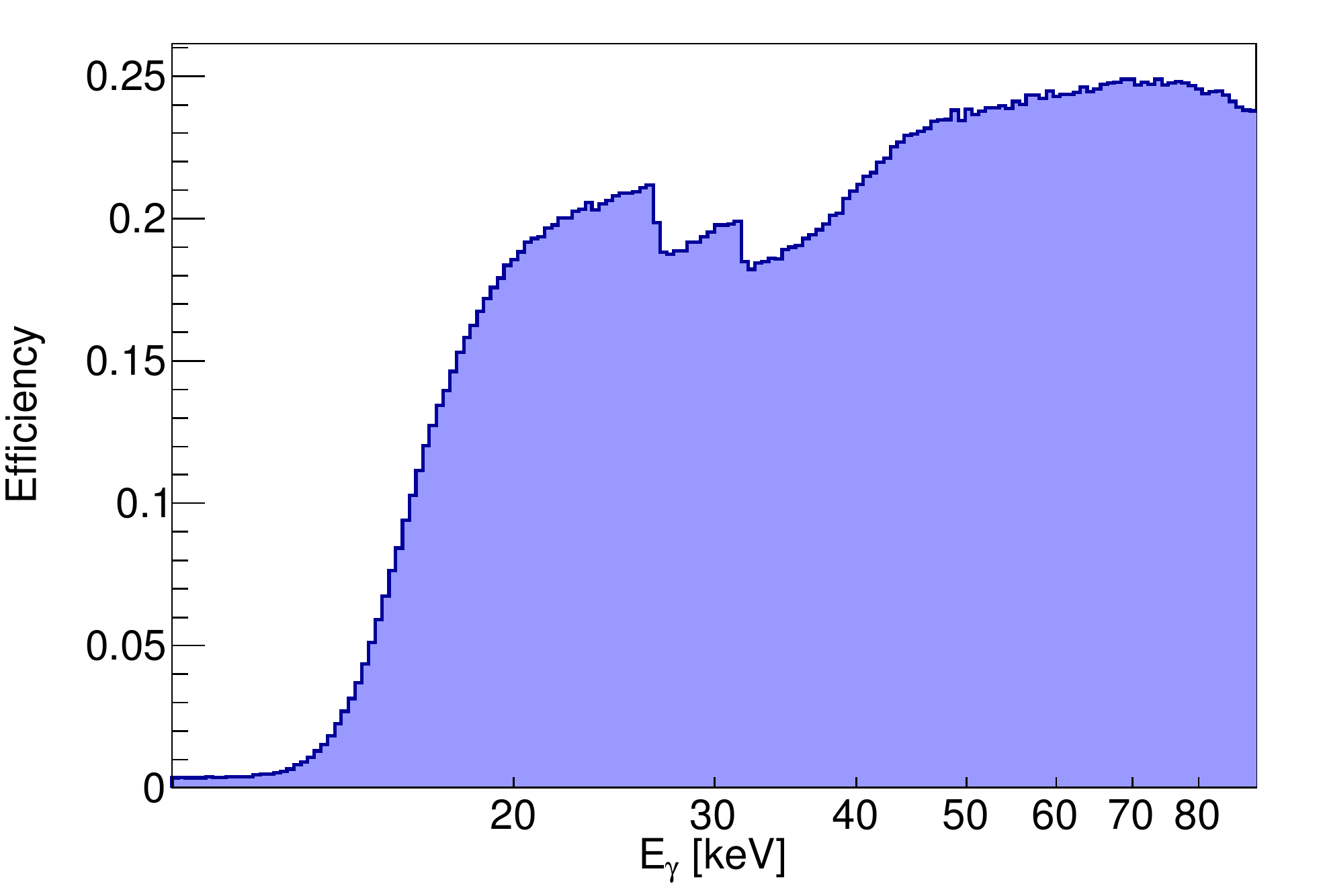}%
  \hfill~%
  \caption{Detection efficiency for photons entering the detector as a function of energy. \textit{Left:} Polarimeter; \textit{right:} \nth{17} CZT. This does not include effects of absorption in the atmosphere and the effective area of the X-ray mirror, which are considered separately. However, all details of the polarimeter, including disabled channels and individual channel thresholds are taken into account.}
  \label{fig:Efficiency}
\end{figure*}

From this simulation we obtained the energy redistribution matrices and detection efficiency curves shown in Figures~\ref{fig:ResponseMatrix} and~\ref{fig:Efficiency}.
The figures do not include effects of the mirror effective area and absorption in the residual atmosphere of the balloon, as those are handled separately in the data analysis.
In particular, the latter changes with time as balloon altitude and pointing elevation vary.
Both figures exclude events in the first row of CZT pixels, which receives some direct, unscattered photons from the X-ray mirror, has a poor signal-to-background ratio, and a low modulation factor, as discussed later.

\begin{figure}
  \centering
  \includegraphics[width=\linewidth]{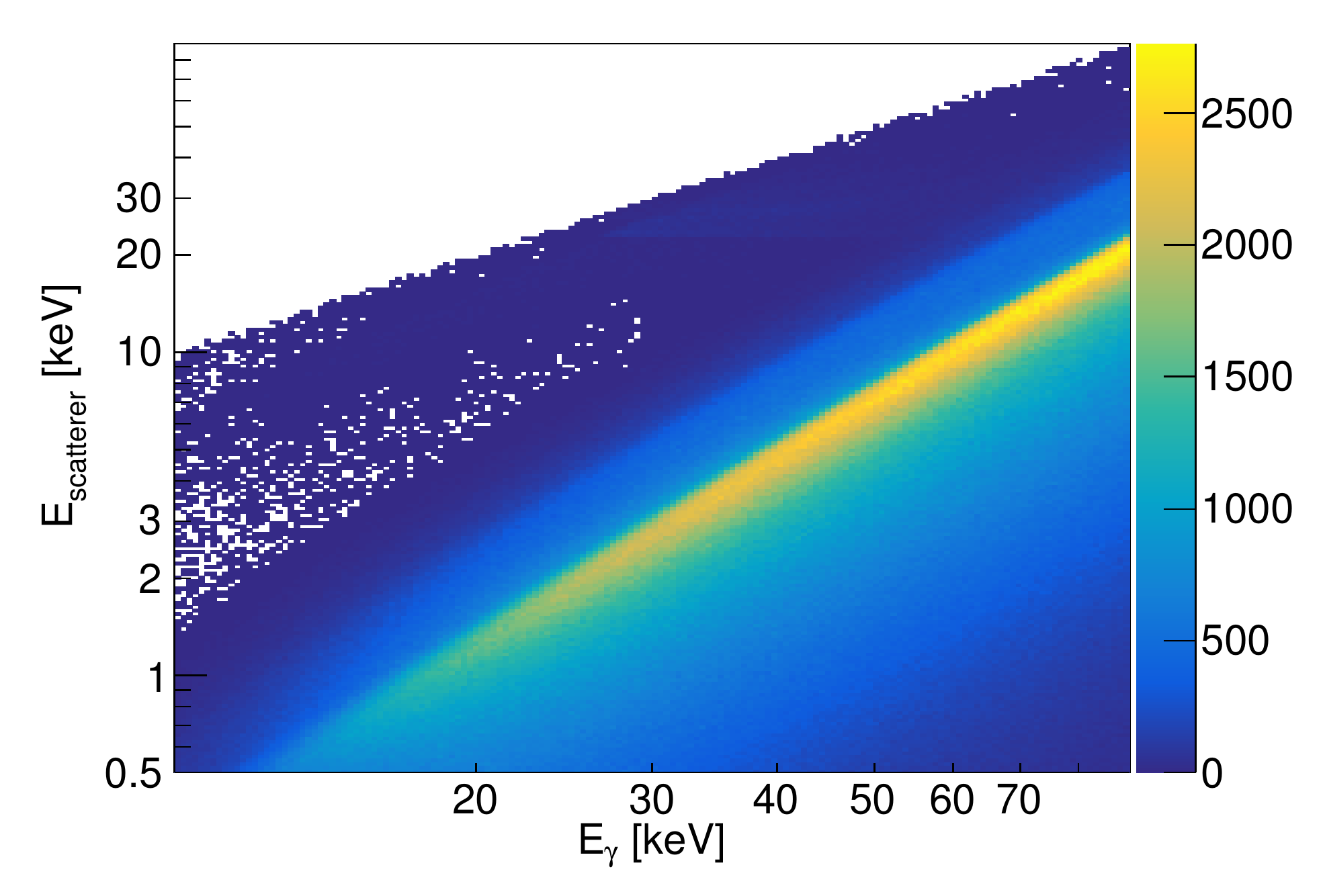}
  \caption{Energy deposition in the scattering element for events that resulted in an energy deposition in a CZT detector as well. This simulation did not consider trigger conditions, and any energy deposition in CZT was accepted.}
  \label{fig:ScattererEnergyDeposit}
\end{figure}

The peak in the response matrix below the main photoabsorption peak is due to fluorescence photons escaping the detector, due to the Cd and Te K$\alpha$, K$\beta$, and K$\gamma$ lines in the \SIrange{22.7}{31.8}{keV} energy range.
The continuum is due to photons Compton scattering in the detectors.
The response of the polarimeter is smoothed by the energy loss when photons Compton scatter in the Be cylinder, which is shown in Fig.~\ref{fig:ScattererEnergyDeposit}.

\begin{figure}
  \centering
  \includegraphics[height=.9\linewidth,angle=270]{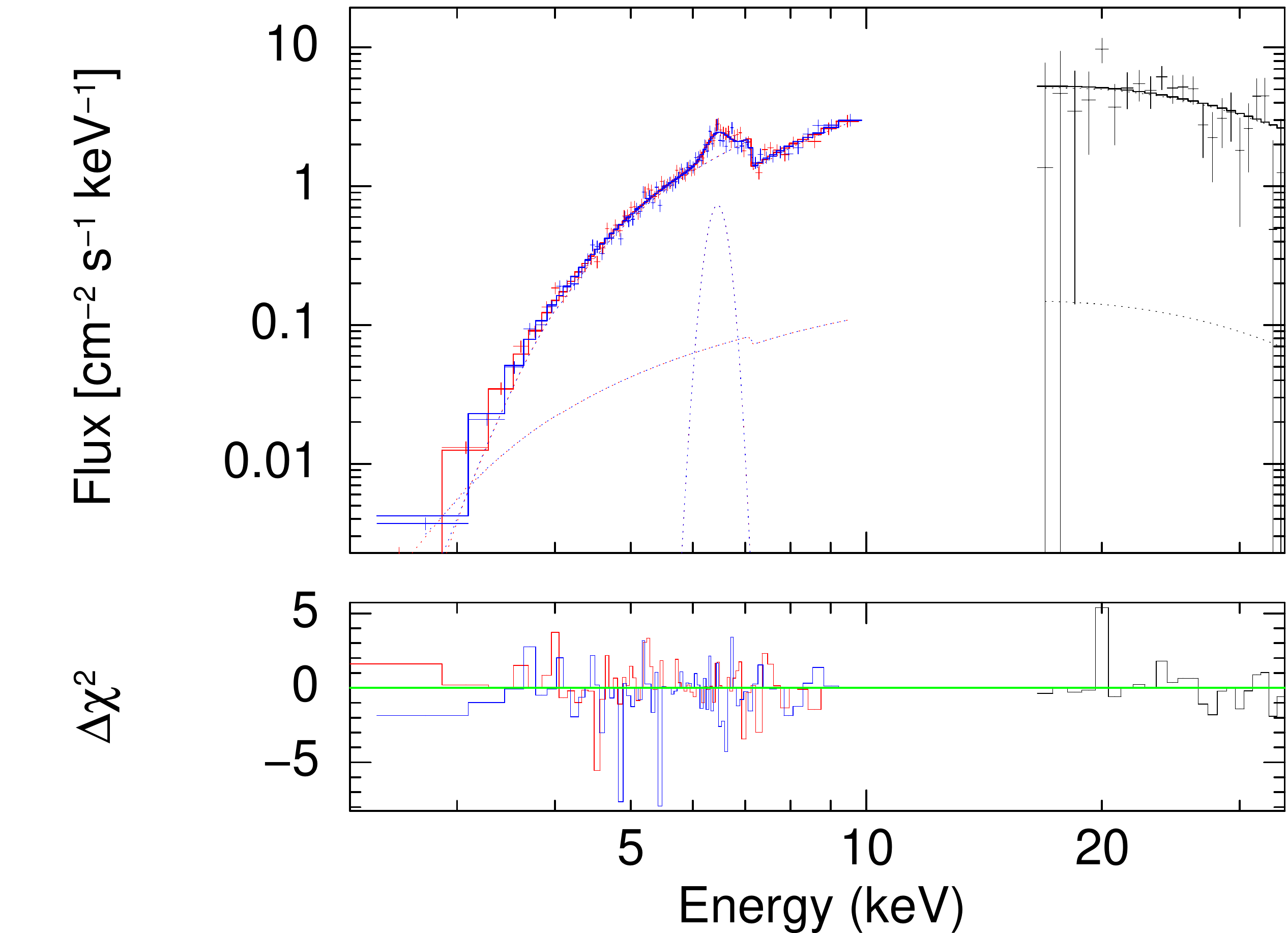}
  \caption{Joint fit of Swift XRT and \xcalibur spectra of GX 301-2 serve as a verification of the effective area simulations of \xcalibur. Reproduced from Ref.~\cite{abarr_etal_2020}.}
  \label{fig:swift_xcal_spectra}
\end{figure}

The minima of the detection efficiency between \SIrange[range-phrase={ and }]{40}{60}{keV} and between \SIrange[range-phrase={ and }]{25}{40}{keV} in the polarimeter and \nth{17} CZT, respectively, are caused by the escape peaks.
In these energy ranges, the energy deposited in the CZT detectors when a fluorescence photon escapes is not sufficient to trigger the CZT detectors.
Due to the thickness of the \xcalibur CZT detectors, the corresponding absorption edges do not result in a significant increase in interaction probability.
Figure~\ref{fig:swift_xcal_spectra} shows a combined fit of Swift XRT and \xcalibur spectra of \mbox{GX 301-2}.
The observations did not overlap in time with the closest observations being separated by about four hours.
We estimate that this allows a verification of the effective area to within a factor~\num{\sim 2}.

\begin{figure}
  \centering
  \includegraphics[width=\linewidth]{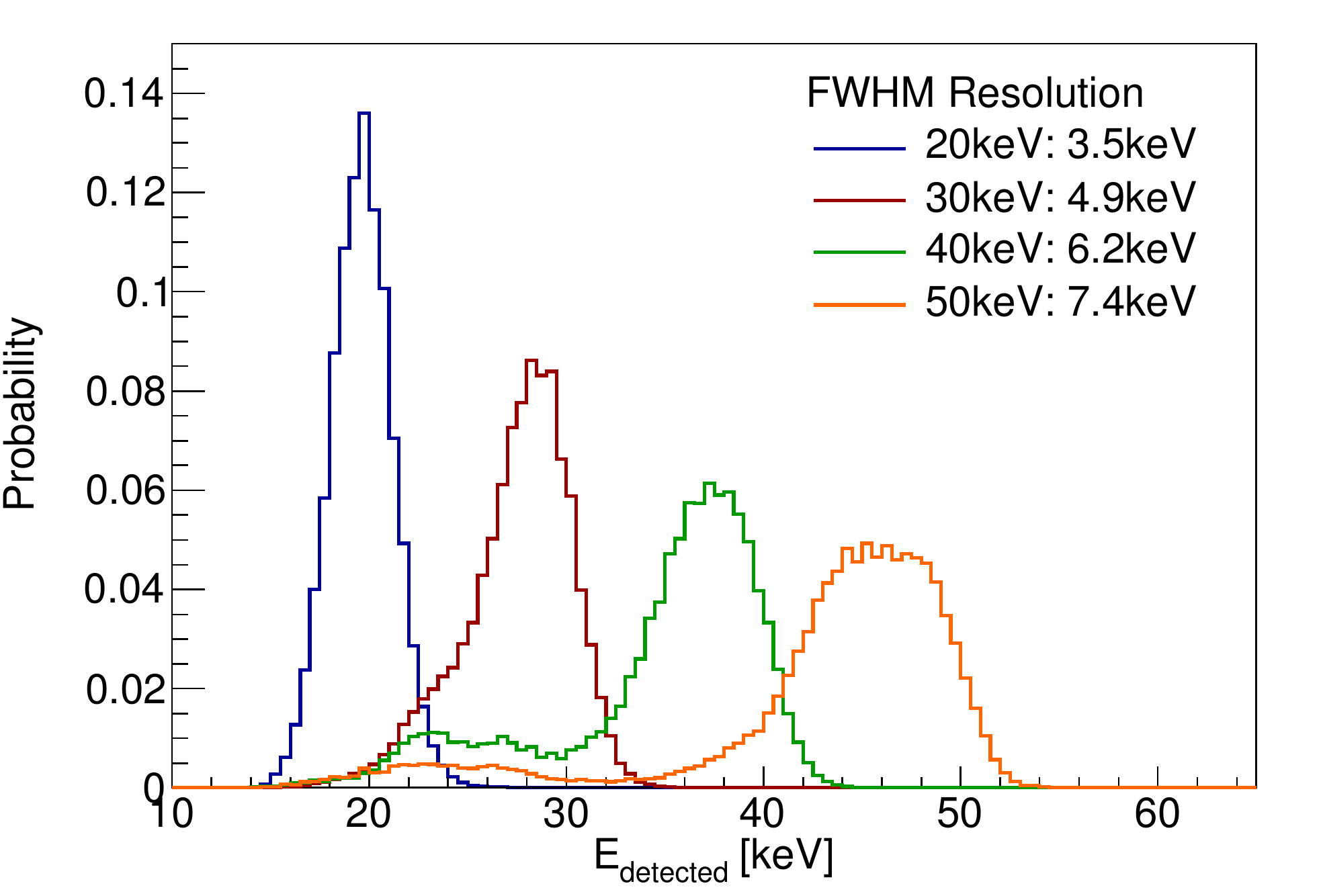}
  \caption{Energy resolution of the polarimeter at a few selected incident photon energies.}
  \label{fig:EnergyResolution}
\end{figure}

Figure~\ref{fig:EnergyResolution} shows the energy resolution of the polarimeter for a few discrete incident photon energies.
As shown in Section~\ref{sub:calibration}, the energy resolution of the CZTs does not depend significantly on incident energy.
The increasing FWHM resolution is due to energy lost in the scattering element as shown in Fig.~\ref{fig:ScattererEnergyDeposit}, which broadens the main photoabsorption peak in the CZT that dominates the energy resolution.
The contribution due to Compton scattering in the CZT detectors and escape peaks is comparatively small.

\begin{figure*}
  \centering
  \includegraphics[width=.47\linewidth]{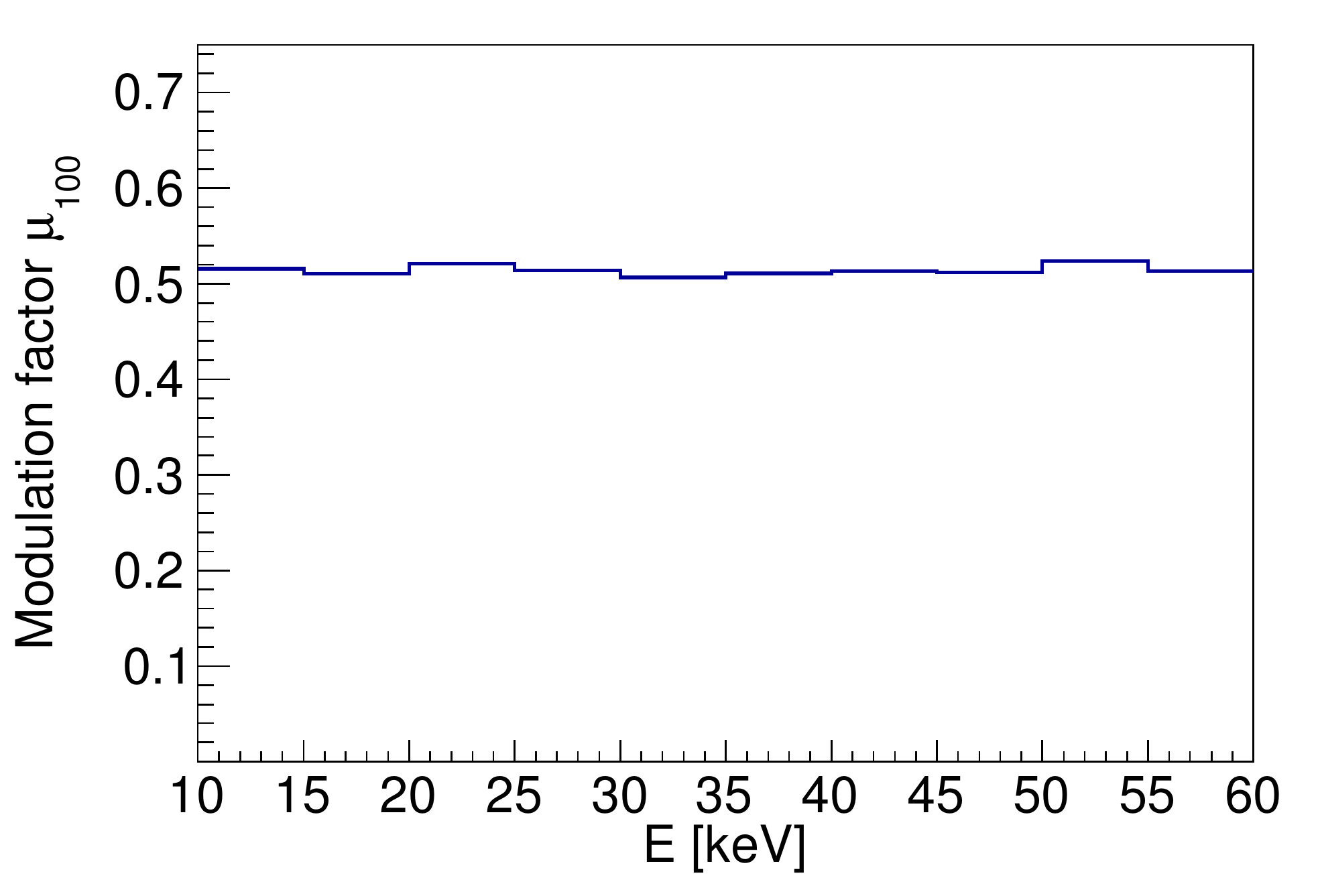}%
  \hfill%
  \includegraphics[width=.47\linewidth]{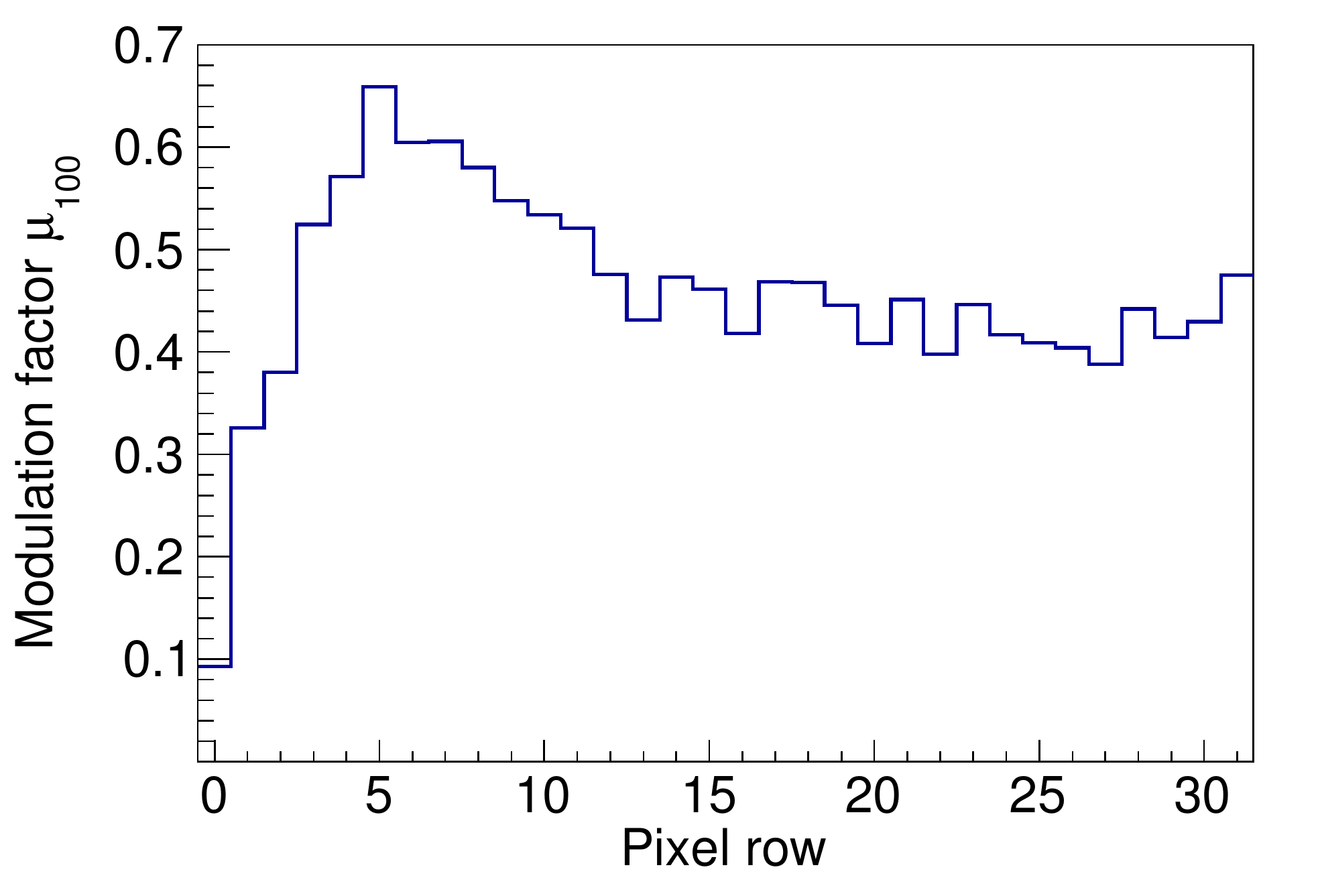}%
  \caption{Modulation factor for all events hitting the polarimeter detector rows \numrange{2}{32} as a function of the energy of the primary photon (left), and as a function of the pixel row for all energies from \SIrange[range-phrase={ to }]{15}{35}{keV} (right). Pixel row 0 is the one closest to the entrance window. The mean modulation factor is $\mu_{100} = \num{0.513}$.}
  \label{fig:mu}
\end{figure*}

Photons scatter preferentially perpendicular to their polarization direction resulting in a sinusoidal modulation of the distribution of azimuthal scattering angles $\phi$:
\begin{equation}
  N(\phi) \propto 1 + p_0\mu_{100}\sin(2(\phi-\Psi_0)),
\end{equation}
where $p_0$ is the polarization fraction, and $\Psi_0$ is the polarization angle.
The modulation factor, $\mu_{100}$, is a parameter of the instrument corresponding to the amplitude of the modulation for a \SI{100}{\percent} polarized beam.
Using simulations of a fully polarized photon beam, we determined the modulation factor of \xcalibur shown in Fig.~\ref{fig:mu}.
Due to the design of \xcalibur, which enables a direct determination of $\phi$ from the position of the hit pixel, $\mu_{100}$ is dictated entirely by the kinematics of Compton scattering.
In the energy range of interest to \xcalibur, there is no significant variation of $\mu_{100}$ with energy.
The right panel of Fig.~\ref{fig:mu} shows the modulation factor as function of pixel row.
Row 0 refers to the pixels closest to the entrance window, and row 31 are the pixels at the rear end of the detector.
The tip of the scattering element is near the center of the first ring of CZT detectors, i.\,e. between pixel rows \num{3} and \num{4}.
The modulation factor peaks just below the tip of the scattering element, where the largest number of photons scattered by \ang{90} are detected.
In this case the suppression of scattering parallel to~$\Psi_0$ is largest.

In addition to the response of the instrument to polarized photons, the 
response to unpolarized photons and the azimuthal distribution of background
events are important as they determine the systematic uncertainties of the
polarization measurement.
The instrument response to an unpolarized beam has been studied in detail
in Ref.~\cite{beilicke_etal_jai_2014}.
For a beam centered on the scattering element, a modulation \SI{<0.8}{\percent} was observed and it was shown that the largest contribution to
asymmetries in the azimuthal response is due to a misalignment of the beam,
which can be corrected for (see Section~\ref{sub:truss}).
Combined with the small residual modulation of the background (see 
Fig.~\ref{fig:background_azimuth}), the overall systematic uncertainty of a
polarization measurement with \xcalibur is estimated to be~\SI{<1}{\percent}.

\section{Summary and Outlook}\label{sec:summary}
\xcalibur is a hard X-ray polarimetry mission covering the \SIrange{15}{50}{keV} energy range.
During its Antarctic flight, it successfully observed GX 301-2 measuring its lightcurve and energy spectrum, as well as providing the first constraints of the hard X-ray polarization of an accretion-powered neutron star.
The instrument combines a focusing X-ray optic with a polarimeter with high detection efficiency, excellent modulation factor \num{>0.5} virtually independent of energy, and good control of systematic errors through continuous rotation of the instrument.
The FWHM energy resolution of the polarimeter ranges from \SI{3.5}{keV} at \SI{20}{keV} to \SI{7.4}{keV} at \SI{50}{keV}.

The first flight demonstrated the viability and sensitivity of using a scattering polarimeter in the focal plane of a pointed x-ray telescope.     
In addition to the scientific results~\cite{abarr_etal_2020}, data collected during the flight of \xcalibur also guide the design of a follow-up mission called \xlcalibur~\cite{abarr_etal_2021}.
\xlcalibur will be significantly more sensitive than \xcalibur due to several improvements:
\begin{itemize}
 \item
  It will utilize a \SI{12}{m} focal-length X-ray mirror with a \numrange{3}{10} times larger effective area.
 \item
  A redesigned balloon gondola and off-axis star tracker will allow pointing up to an elevation of \ang{\sim 82}, compared to a limit of \ang{65} imposed by the \xcalibur gondola and the fact that the on-axis star tracker does not allow pointing through the balloon (which is transparent to X-rays).
  This will allow us to observe high-elevation sources when absorption in the atmosphere is the lowest.
  \item The use of the lower-noise NRL-2 ASICs will lower the trigger threshold of the CZT detectors to $\sim$12\,keV, safely below the 
  lowest energy of 15\,keV at which the rest-atmosphere becomes 
  transparent to X-rays. In comparison, roughly half of the 
  {\it XL-Calibur} detector channels had energy thresholds between 
  15\,keV and 20\,keV (Fig.\,\ref{fig:thresholds}), 
  leading to a non-negligible sensitivity loss.
 \item
  CZT detectors with a thickness of \SI{0.8}{mm} will achieve slightly better energy resolutions and will reduce the background by a factor \num{1.8} compared to \xcalibur's \SI{2}{mm} thick detectors.
\item
  The CsI(Na) anti-coincidence shield will be replaced by a new BGO shield with a higher stopping power, and the polarimeter readout boards will be redesigned to allow a more compact assembly.
  The shield readout electronics will be improved based on the results presented in Section~\ref{sub:shield} in order to reduce the veto threshold to \SI{100}{keV} while keeping the deadtime low.
  Simulations similar to those presented in Section~\ref{sec:background} show that these changes will reduce the background rate by a factor \numrange{10}{25}, allowing observation of much fainter objects.
\end{itemize}

\xcalibur has pioneered and validated a new approach to hard X-ray polarimetry.
\xlcalibur will build on this heritage and provide highly significant polarization measurements of several sources in the \SIrange{15}{80}{keV} energy range.
It will complement polarization measurements in the \mbox{\SIrange{2}{8}{keV}} energy range by IXPE, which was successfully launched on December 9th 2021~\cite{weisskopf_2016a,weisskopf_2016b,weisskopf_2018}.

An \xcalibur-type polarimeter can readily be adapted for use on a Small or Medium Explorer (SMEX or MIDEX) mission~\cite{krawczynski_etal_2016}.
As a standalone mission, such an instrument could cover the \SIrange{3}{80}{keV} energy range.
An ideal mission would combine the hard X-ray polarimeter with a REDSOX-type~\cite{marshall_etal_2018} soft X-ray polarimeter and an instrument similar to IXPE or PRAXyS~\cite{iwakiri_etal_2016} for the intermediate energy range.
Such a mission has been proposed~\cite{krawczynski_etal_2019,jahoda_etal_2019} to enable simultaneous measurements of X-ray polarization from a few hundred \si{eV} up to \SI{\sim80}{keV}.
It would allow powerful precision tests of emission models by simultaneously observing several emission components.

\section*{Declaration of Competing Interests}
The authors declare that they have no known competing financial interests or personal relationships that could have appeared to influence the work reported in this paper.

\section*{Acknowledgements}
\xcalibur is funded by the NASA APRA program under contract number 80NSSC18K0264.
We thank the McDonnell Center for the Space Sciences at Washington University in St.~Louis for funding of an early polarimeter prototype, as well as for funds for the development of the ASIC readout.
H.K.~acknowledges NASA support under grants 80NSSC18K0264 and NNX16AC42G.
KTH authors acknowledge support from the Swedish National Space Agency (grant No. 199/18).
M.P.~also acknowledges support from the Swedish Research Council (grant No. 2016-04929).
H.K.~acknowledges support from the National Science Foundation under the Independent Research and Development program.
We thank Rakhee Kushwah (KTH, Oskar Klein Centre) for contributing to the flight monitoring shifts.

\sloppy
\bibliographystyle{elsarticle-num-abbrv}
\bibliography{xcalibur}

\end{document}